\documentclass[12pt,preprint]{aastex}
\usepackage{graphicx}
\input ./insertplot.ps

\begin{document}

\title{{\em Hubble Space Telescope} ACS Imaging of the GOALS Sample: Quantitative Structural Properties
of Nearby Luminous Infrared Galaxies with $L_{\rm IR} > 10^{11.4}$ L$_\odot$
 \footnote{Based on observations with the NASA/ESA {\em Hubble Space Telescope},
  obtained at the Space Telescope Science Institute, which is operated
  by the Association of Universities for Research in Astronomy,
  Inc. under NASA contract No.  NAS5-26555.}}

\author{
D.-C. Kim\altaffilmark{1,2},
A. S. Evans\altaffilmark{1,2},
T. Vavilkin\altaffilmark{3},
L. Armus\altaffilmark{4},
J. M. Mazzarella\altaffilmark{5},
K. Sheth\altaffilmark{2},
J. A. Surace\altaffilmark{4},
S. Haan\altaffilmark{4,*},
J. H. Howell\altaffilmark{4},
T.~D\'{\i}az-Santos\altaffilmark{5},
A. Petric\altaffilmark{6},
K. Iwasawa\altaffilmark{7},
G. C. Privon\altaffilmark{1},
\& D. B. Sanders\altaffilmark{8}
}
\altaffiltext{1}{Department of Astronomy, University of Virginia, 530
McCormick Road, Charlottesville, VA 22904: dkim@nrao.edu; aevans@virginia.edu; gcp8y@virginia.edu}

\altaffiltext{2}{National Radio Astronomy Observatory, 520 Edgemont Road,
Charlottesville, VA 22903: astrokartik@gmail.com}

\altaffiltext{3}{Department of Physics \& Astronomy, Stony Brook
University, Stony Brook, NY, 11794-3800: tvavilkin@googlemail.com}

\altaffiltext{4}{Spitzer Science Center, Pasadena, CA 91125:
lee@ipac.caltech.edu; jason@ipac.caltech.edu; jhhowell@ipac.caltech.edu}

\altaffiltext{*}{Current Address: CSIRO - Astronomy \& Space Science, PO Box 76,
Epping NSW 1710, Australia: sebastian.haan@csiro.au}

\altaffiltext{5}{Infrared Processing and Analysis Center, California
Institute of Technology, MS 100-22, Pasadena, CA 91125: mazz@ipac.caltech.edu; tanio@ipac.caltech.edu}

\altaffiltext{6}{Astronomy Department, California
Institute of Technology, MC 249-17, Pasadena, CA 91125: ap@astro.caltech.edu}

\altaffiltext{7}{ICREA and Institut de Ci\`encies del Cosmos (ICC), Universitat de
Barcelona (IEEC-UB), Mart\'i i Franqu\`es, 1, 08028 Barcelona, Spain: kazushi.iwasawa@icc.ub.edu}

\altaffiltext{8}{Institute of Astronomy, University of Hawaii, 2680
Woodlawn Dr., Honolulu, HI 96822: sanders@ifa.hawaii.edu}

\begin{abstract}
A {\it Hubble Space Telescope} ({\it HST}) / Advanced Camera for Surveys (ACS) 
study of the structural properties of 85 
luminous and ultraluminous ($L_{\rm IR} > 10^{11.4}$ L$_\odot$) infrared galaxies (LIRGs and ULIRGs) 
in the Great Observatories All-sky LIRG Survey (GOALS) sample is presented.
Two-dimensional GALFIT analysis has been performed on F814W ``{\it I}-band'' images to decompose each galaxy, as appropriate, into bulge, disk, central PSF and stellar bar components.
The fraction of bulge-less disk systems is observed to be higher in LIRGs (35\%) than in ULIRGs (20\%), with the disk+bulge systems making up the dominant
fraction of both LIRGs (55\%) and ULIRGs (45\%). Further, bulge+disk systems are the dominant late-stage merger galaxy
type and are the dominant type for LIRGs and ULIRGs
at almost every stage of galaxy-galaxy nuclear separation.
The mean {\it I}-band host absolute magnitude of the GOALS galaxies is $-22.64\pm$0.62 mag (1.8$^{+1.4}_{-0.4}$ L$^*_I$),
and the
mean bulge absolute 
magnitude in GOALS galaxies is about 1.1 magnitude fainter than the mean host magnitude. Almost all
ULIRGs have bulge magnitudes at the high end ($-20.6$ to $- 23.5$ mag) of the GOALS bulge magnitude range.
Mass ratios in the GOALS binary systems are consistent with most of the galaxies being the result of major mergers, and
an examination of the residual-to-host intensity ratios in GOALS binary systems suggests that
smaller companions suffer more tidal distortion than the larger companions.
We find approximately twice as many bars in GOALS disk+bulge systems (32.8\%) than in pure-disk mergers (15.9\%) but most of the disk+bulge systems that contain bars are disk-dominated with small bulges. 
The bar-to-host intensity ratio, bar half-light radius, and bar ellipticity in GOALS galaxies
are similar to those found in nearby spiral galaxies.
The fraction of stellar bars decreases toward later merger stages and smaller nuclear separations, indicating that bars are destroyed as the merger advances.
In contrast, the 
fraction of nuclear PSFs increases towards later merger stages and is highest in late-stage systems with a 
single nucleus. Thus, light from an AGN or compact  nuclear star cluster is more visible at I-band as ULIRGs enter their
latter stages of evolution.
Finally, both GOALS elliptical hosts and nearby SDSS ellipticals occupy the same part of the surface brightness vs. half-light radius plot (i.e., the ``Kormendy Relation'') and have similar slopes,
consistent with the possibility that the GOALS galaxies belong to the same parent population as the SDSS ellipticals. 
\end{abstract}


\section{Introduction}

The Great Observatories All-sky LIRG Survey (GOALS) combines 
imaging and spectroscopic data from NASA's {\it Spitzer}, {\it HST}, {\it Chandra}
and {\it GALEX} space-borne observatories in a comprehensive study of
the most luminous infrared-selected galaxies in the local Universe (Armus et al. 2009).
The sample consists of 181 Luminous Infrared Galaxies (LIRGs, $L_{\rm IR} = 10^{11.0} - 10^{11.99}$L$_\odot$)
and 21 ultraluminous infrared galaxies (ULIRGs, $L_{\rm IR} \ge 10^{12}$L$_\odot$) derived from the IRAS Revised Bright Galaxy Sample (RBGS, Sanders et al. 2003).
The sample spans the full range of optical
nuclear spectral types (starbursts, LINERs, type-1 and type-2 Seyferts; i.e., Veilleux et al. 1995) as well as interaction stages, 
and serves as a statistically complete sample of infrared-luminous, local galaxies. As such, the GOALS galaxies
are excellent analogs for comparisons with infrared and submillimeter-selected galaxies at high redshift.

The Digitized Sky Survey (DSS) images of the RBGS (U)LIRGs at $L_{\rm IR} > 10^{11.0}$L$_\odot$ (Sanders et al. 2003) show that more than 90\% of them have signs of tidal interaction.
The galaxy interaction efficiently drives gas to the central regions where vigorous starburst activity 
(Barnes \& Hernquist 1992), and possibly the accretion of gas unto supermassive nuclear
black holes, can commence.
Galactic outflows have been found in many (U)LIRGs 
(Armus et al. 1989; 1990, Heckman et al. 1990, Rupke et al. 2005b, Sturm et al. 2011, Chung et al. 2011). 
The galactic scale outflows from stellar winds and supernovae explosions play an important role in galaxy evolution.
The galactic superwinds clear out nuclear gas and dust, enrich the intergalactic medium, 
and eventually quench star formation activity and blackhole growth, 
and turn gas-rich spiral progenitors into elliptical galaxies or S0-type galaxies (Barnes 2002, Veilleux et al. 2005, Sturm et al. 2011)

Recently, we have undertaken morphological studies of 73 GOALS (U)LIRGs with the HST NICMOS H-band images (Haan et al. 2011, hereafter Paper I).
We find that a significant fraction of the GOALS (U)LIRGs have double (63\%) or triple nuclei (6\%) 
which were not seen in the B and I band HST data due to obscuration, 
and the bulge luminosity surface density increases significantly along the merger sequence, 
while the bulge luminosity shows a small increase toward late merger stages.
This increase in the luminosity surface density was 
found by Haan et al. to be almost entirely due to a decrease of the bulge radius 
in the late stage merging (U)LIRGs, 
consistent with an inside-out growth of the bulge due to the funneling of gas towards the centers.
Haan et al. also found
that the projected nuclear separation is significantly smaller for ULIRGs (median value of 1.2 kpc) than for LIRGs
(median value of 6.7 kpc), suggesting that the LIRG phase appears at an earlier merger stage than the ULIRG phase.
The GOALS H-band images are better suited for finding multiple nuclei, identifying embedded star clusters and measuring the structural parameters in the nuclear region
in these dusty GOALS galaxies, but the fields of view are generally too small 
to study the large scale morphologies, bulge-to-disk ratios, tidal features, and stellar bars, 
except in the more distant sources.
To address these latter issues, we have undertaken morphological studies of the GOALS (U)LIRGs with the HST ACS I-band images.  Optical images remain excellent tools for measuring bar fractions and bar length, as long as the rest-frame wavelengths are long wards of the Balmer break  (e.g., Sheth et al. 2003, 2008; Menendez-Delmestre et al. 2007, 2012).  
The present paper aims to study optical structural properties of GOALS (U)LIRGs with $L_{\rm IR} \geq 10^{11.4}$ L$_\odot$ using GALFIT (Peng et al. 2002, 2010).

The ACS on {\it HST} provides superior spatial resolution, making it possible to 
image detailed structural features
such as unresolved nuclei, faint tails, bridges, rings, and shells. In addition, the large field of view ($202\arcsec \times 202\arcsec$) of the Wide Field Channel (WFC) on ACS
makes it possible to image (U)LIRGs consisting of widely separated pairs in a single orbit. The structural decomposition
of GOALS (U)LIRGs allow an assessment of the relative fractions of (U)LIRGs with disks, disks+bulges, and elliptical-like profiles, 
the structural properties of the progenitors (early stage mergers) and evolutionary byproducts (late-stage mergers) of
LIRGs and ULIRGs, and their fundamental plane properties.  
Further, the present sample spans all the merger evolutionary stages (e.g., Sanders et al. 2003), making it 
possible to assess these properties as a function of merger phase.
Note, however, that our sample was selected, in part, by luminosity cut ($L_{\rm IR} \geq 10^{11.4}$ L$_\odot$)
and thus may be biased toward later stage mergers. The results presented here are representative of all (U)LIRGs with $L_{\rm IR} \geq 10^{11.4}$ L$_\odot$ in the local universe.

The paper is divided into five Sections. In \S 2, the sample selection and observations are summarized. 
In \S3, the data reduction and analysis are described. 
The results and discussion are presented in \S 4 \& \S 5, 
and a summary of the paper is presented in \S 6.
Throughout this paper, the cosmology $H_0$ = 70 km s$^{-1}$
Mpc$^{-1}$, $\Omega_M$ = 0.3, and $\Omega_\Lambda$ = 0.7 are adopted (see also Armus et al. 2009).

\section{Sample Selection and Observations}

Sample galaxies were selected from a complete sample of 87 (U)LIRGs with $L_{\rm IR} \geq 10^{11.4}$ L$_\odot$ 
in the IRAS Revised Bright Galaxy Sample (RBGS, i.e., $f_{60} > 5.24$ Jy and Galactic Latitude $|b| > 5^{\rm o}$, Sanders et al. 2003). 
These (U)LIRGs have been imaged with {\it HST} with the ACS/WFC
using the F435W (B) and F814W (I) broad-band filters (GO program 10592, PI: A. Evans: see Evans et al. 2013).
One galaxy was observed per orbit in the ACCUM mode, with total exposure times of $\sim 21$ minutes and $\sim 12$ minutes in F435W and F814W filters, respectively.
The F435W and F814W observations were done using the three and two point line dither patterns, respectively.
Further details of the observations can be found in Evans et al. (2013).
The basic properties of GOALS galaxies in our sample are listed in Table 1.

\section{Data Reduction and Analysis}

The data reduction was initiated with the flat-fielded science file 
retrieved from the Multimission Archive at Space Telescope (MAST).
The four quadrants of the flat-fielded images showed a varying range of bias level offsets (up to 20\%); these quadrant offsets were measured and corrected, and the sky background was set to zero.
Next, the cosmic-rays in the images were removed with the lacosmic routine (van Dokkum 2001), which 
uses a Laplacian edge detection algorithm. The algorithm works better on non-drizzled images since drizzling smooths the sharp edges.
Individual images with the cosmic-ray removed were combined using Multidrizzle package (Koekemoer et al. 2002); this routine
also removes additional cosmic-rays and corrects the geometric distortion of the images.
The remaining cosmic rays were further processed with jcrrej2 routine in IRAF (Rhoads 2000).
Finally, the combined images were rotated and WCS header informations were corrected by making use of 2MASS catalog
coordinates of bright stars in each image.

Model fitting of the images was done using the 2-dimensional galaxy fitting package GALFIT. 
The procedure was used to 
fit the central point source in each object and determine the structural parameters of the underlying host galaxies. 
The analysis of each object followed a number of well defined steps. 
First, a mask was constructed to exclude bright stars or small foreground/background galaxies within the field of view.
Next, because we are particularily interested in disk and bulge properties in underlying host galaxies,
a fit was made to the surface brightness profile of each object for the following cases of the disk and bulge combinations:

\noindent
{\it (i)} PSF + exponential disk profile ($n = 1$) alone,\\ 
{\it (ii)} PSF + de Vaucouleurs profile ($n = 4$) alone, and\\
{\it (iii)} PSF + disk ($n = 1$) and bulge ($n = 4$),

\noindent
where PSF is the point-spread-function and $n$ is the power-law index of the Sersic profile (Sersic 1963), i.e.,

$$\Sigma (r) = \Sigma _e e^{-\kappa [(r / r_e)^{1/n} - 1]}. \eqno(1)$$

\noindent
In Equation (1), $r_e$ and $\Sigma _e$ are the effective radius and surface brightness, respectively.
In cases where a bar can visually be identified in a disk and disk+bulge galaxies (i.e., 25.9\% of the sample), a bar component was included in the 2-D fit.
Next, chi-square values of the fit, fitted components, and residual image for the above 3 cases were examined to identify the best fit model among these 3 cases.
For the galaxies with multiple nuclei, the fitting was done simultaneously.

Early in the fitting process, 
it was determined that the fits to the F435W images did not converge; the dusty nature of these
(U)LIRGs renders their appearance at 0.4$\mu$m too patchy and discontinuous for a GALFIT analysis. These fits were thus abandoned -- only
fits to the F814W data are included and discussed here. Note also that nuclei of two of the 87 (U)LIRGs observed with {\it HST},
i.e., Mrk 231 and IRAS 05223+1908, are saturated at F814W, and thus reliable fits for these (U)LIRGs could not be
achieved.
Fig. 1 contains an example image of the galaxy, the model version of the galaxy, the residual (i.e., galaxy - model fit), and the
individual model components (PSF, bulge, disk, and bar).
Table 2 lists structural parameters derived from the best fit model for each of the remaining 85 (U)LIRGs (total number of nuclei is 137: 38 singles + 42 doubles + 5 triples) 
in the F814W filter.

\section{Results}

As stated earlier, the sample contains wide range of interaction classes from
advanced mergers to widely separated systems with nuclear separation up to 65 kpc.
In this sense, the GOALS sample represents an evolutionary sequence for interacting galaxies.
The current GOALS sample consists of 64 LIRGs (75.3\%) and 21 ULIRGs (24.7\%) and
will be discussed in terms of these luminosity classes where appropriate.
For the objects with multiple nuclei, the infrared luminosity for each nucleus will be apportioned according to 
its Wide-field Infrared Survey Explorer (WISE) 22$\mu$m flux or Two Micron All Sky Survey (2MASS) K-band magnitude.
Throughout the paper, the term ``single-nucleus merger (U)LIRGs'' is used to refer to (U)LIRGs in which the nuclei
of their progenitors have coalesced and thus a single nucleus is observed. 

\subsection{Galaxy Morphological Types}

The fractions of morphological types derived from the GALFIT analysis for the GOALS host galaxies are listed
in Table 3 and plotted in Fig. 2.
The fraction of disk, disk+bulge, and elliptical in GOALS galaxies are 32.1\%, 48.9\%, and 19.0\% respectively.
If we subdivide GOALS sample into LIRGs and ULIRGs, the distributions of morphological types are rather different.
Both classes have a similar fraction of disk+bulge galaxies, but a higher fraction of disks are found in GOALS LIRGs,
whereas a higher fraction of ellipticals are found in GOALS ULIRGs.
The mean (range) projected nuclear separations of the GOALS LIRGs and GOALS ULIRGs are 
12.2 kpc (0 - 65.1 kpc) and 4.0 kpc (0 - 39.2 kpc) respectively.
This suggests that more luminous, advanced merger systems tend to have a relatively smaller disk fraction and a higher elliptical fraction. Note, 
however, that the disk+bulge galaxy types make up the higher fraction of both LIRGs (55\%) and ULIRGs (45\%).

The distribution of morphological type as a function of interaction class (see Haan et al. 2011; Evans et al. 2013 for more details) is plotted in Fig. 3.
The following is a brief description of interaction class (IC) based solely on a visual inspection of each system.\footnote{
More detailed modeling of the interaction stages which takes into account kinematics is underway (Privon, Ph. D. Thesis, Privon et al. 2013).
This modeling will allow us to address obvious degeneracies in our present classification scheme.}

\indent
  0 - single undisturbed galaxy, shows no signs of tidal interaction\\
\indent
  1 - separate galaxies, disks symmetric (intact), no tails\\
\indent
  2 - progenitor galaxies distinguishable, disks asymmetric or amorphous, tidal tails\\
\indent
  3 - two nuclei in a common envelope\\
\indent
  4 - double nuclei + tidal tail\\
\indent
  5 - single or obscured nucleus, long prominent tails\\
\indent
  6 - single or obscured nucleus, disturbed central morphology, short faint tails or tails absent, shells.

At the widely separated stage (IC=1), elliptical hosts occupy the smallest fraction of morphological types (11\%) 
whereas disk$+$bulge systems occupy the largest fraction (61\%).
The fraction of disk galaxies gradually increases from IC=1 to IC=3 and decreases from IC=3 to IC=5.
In contrast, the disk$+$bulge fraction decreases from IC=1 to IC=3 and increases from IC=3 to IC=6, and, 
with the exception of IC = 3, they make up the highest fraction of (U)LIRGs at every interaction stage.
In general, there is no clear trend in morphological types as a function interaction class,
but we find more elliptical fraction in advanced single-nucleus merger (IC=5) than in widely separated galaxies (IC=1).
The distribution of morphological types in isolated systems (IC=0) is similar to that in advanced mergers (IC=5 \& IC=6)
and all but one galaxy in IC=0 are either disk+bulge or elliptical galaxies.
This may be due to the fact that many IC=0 systems are late-stage mergers in which the tidal features have faded.

\subsection{Host Properties}

Fig. 4 shows distributions of host (total - PSF component, left panel) and bulge (right panel) absolute magnitudes for GOALS galaxies (red),
GOALS LIRGs (green), and GOALS ULIRGs (blue) respectively.
The vertical dashed line represents $M^*_I=-22.0$ mag (Blanton et al. 2003), i.e., the $I$-band absolute magnitude of an
$L^*$ galaxy in a Schechter luminosity function of the local galaxies. 
The host absolute magnitudes (for double and triple galaxies, all components were summed 
since these interacting components will eventually be completely merged to form a single nucleus system) 
for the GOALS galaxies range from $-20.98$ to $-23.94$ magnitudes, with a mean (median) of $-22.64\pm$0.62 ($-$22.75) mag.
There is essentially no difference in the mean absolute magnitudes between the GOALS LIRGs and GOALS ULIRGs 
($-22.61\pm$0.68 mag for LIRGs and $-22.72\pm$0.41 mag for ULIRGs).
The average magnitude of the GOALS host galaxies corresponds to 1.8$^{+1.4}_{-0.4}$ L$^*_I$.

The bulge absolute magnitudes of the disk+bulge and elliptical hosts range from $-17.54$ to $-23.70$ mag 
with a mean (median) of $-21.56\pm$1.03 ($-21.77$) mag.
The mean bulge luminosity is about 1.1 mag fainter than the mean host galaxy luminosity.
Although the distributions are different, we do not find a significant difference 
of mean bulge absolute magnitude between the GOALS LIRGs and GOALS ULIRGs ($-21.38\pm$1.06 and $-22.07\pm$0.78 for LIRGs and ULIRGs respectively).

The distribution of bulge-to-disk ratio, B/D, is plotted in Fig. 5. 
The B/D ratios (i.e. ratio of Sersic index of n=4 bulge component to n=1 disk component from GALFIT) range from 0.01 to 7.9 with a mean (median) of 1.23$\pm$1.66 (0.67).
Most of the B/D in GOALS galaxies are less than 2, and more than 60\% are less than 1, 
thus the majority of the GOALS galaxies are disk-dominated system.
The median B/D value of 0.67 in GOALS galaxies is comparable to that of the local 
Sa-Sab galaxies (Oohama et al. 2009).

In binary systems, a mass ratio between two interacting galaxies can be estimated from their luminosity ratio by assuming that they have the same M/L ratio.
Fig. 6 shows the distribution of mass ratio in the binary systems in interacting galaxies (Fig. 6a) and
the mass ratio as a function of M(host) (Fig. 6b).
The result suggests that most of the GOALS binaries are major mergers (i.e., mass ratio of $m_{secondary}/m_{primary}>$1/4) and only 7\% are minor mergers.
The mean value of the mass ratio is 0.55$\pm$0.23 in all GOALS binaries.
The mean mass ratio of the GOALS ULIRGs (0.60$\pm$0.23) is similar to that of the GOALS LIRGs (0.55$\pm$0.23).
In general, the absolute magnitude of the equal mass interacting galaxies
is brighter than that of the unequal mass interacting galaxies.
However, this correlation is very weak (r=0.34) as shown in M(host) and mass ratio plot (Fig. 6b).

\subsection{Stellar Bars}

Bars are ubiquitous in disk galaxies and play an important role in the dynamical and secular evolution of galaxies. A series of GALFIT measurements with a bar component included was run on the $I$-band images of the GOALS sample.
GALFIT successfully decomposed bars for the isolated and mildly disturbed systems,
however, such a decomposition was unsuccessful for very disturbed systems.
For these disturbed systems, we have relied on visual inspections to identify the bars.
The result of bar decomposition is listed in Table 2: if the bars were identified bar by visual inspection, it is flagged as `Y' in column 9 and if not, it is flagged as 'N'.
If the bar components were successfully fitted with GALFIT, it is flagged as `B' in column 2.

\subsubsection {Bar fraction}

The bar fraction in nearby, bright spiral galaxies is $\sim$65\% (de Vaucouleurs 1963) with a third classified as strong (SB) bars and another third classified as intermediate (SAB) bars.  A variety of studies have confirmed this fraction (e.g., Moles et al. 1995; Mulchaey \& Regan 1997; Hunt \& Malkan 1999; Eskridge et al. 2000; Knapen et al. 2000; Laurikainen et al. 2004; Sheth et al. 2008).  Studies in the infrared, where the stellar bars are more easily identified, have also confirmed this bar fraction and have shown that many of the "intermediate" (SAB) bars in the optical are re-classified as "strong" (SB) bars in the infrared (e.g. Eskridge et al. 2000; Menendez-Delmestre et al. 2007; Sheth et al. 2012) but the overall bar fraction in bright, nearby disk galaxies is unchanged.  At the same time, some studies which have quoted lower bar fractions in spirals (e.g. 48--55\%,  Aguerri et al. 2009 or Barazza et al. 2008)  can be attributed to different, and less accurate bar identification techniques or different sample selection criteria.  In comparison to these largely non-interacting, normal disk galaxies, the overall bar fraction in the GOALS sample is much lower (25.9\% - 29 out of 112 galaxies).

The bar fraction as a function of interaction class and morphological type are presented 
in Fig. 7 (fraction on top of each histogram represents number of galaxies with bar to the total number of galaxies in the bin)
and tabulated in Table 4.
The bar fraction gradually decreases from widely separated systems to more advanced merger systems.
When galaxies are relatively well separated (IC=1), about 56\% of the galaxies have bars.
This fraction is similar to that in nearby spiral galaxies.
However, if the galaxy orientation is considered (see below), this bar fraction represents a lower limit.
In most cases, the studies of bar fraction have been restricted to galaxies having inclination angles less than $i<60\sim70^{\circ}$.  If the GOALS galaxies are randomly orientated and a reliable bar decomposition can only be applied to galaxies with orientation angle $< 70^{\circ}$,
then the bar fraction of 56\% found in well separated galaxies (IC=1, mean nuclear separation=$\sim 40$ kpc) would effectively be boosted to 72\% (9/7$\times$56\%).
This fraction is about 10\% to 20\% more than that found in optical images of spirals; if the simplistic extrapolation is correct, it is suggestive of
tidally-induced bar formation (Toomre \& Toomre 1972; Noguchi 1987; Gerin et al. 1990; Steinmetz \& Navarro 2002) already commencing at nuclear separation of $\sim 40$ kpc.

The mean bar fraction in completely merged systems (IC=5 and IC=6) is much smaller than that of the whole sample suggesting possible destruction of bars at the terminal stage of the interaction.
The bar fraction in IC=0 is 28.6\% which is approximately  5 times larger than that of the single nucleus mergers (IC=5 and IC=6) but about a half of that found in nearby spirals.
In Fig. 7b, the bar fraction in GOALS disk+bulge systems (32.8\%) is about twice that observed in disk galaxies (15.9\%).
If we calculate B/D ratios in bulge+disk systems that contain bar, most of them are disk-dominated systems (B/D $<$1) and  their median value of B/D is only 0.18.
This value is about 1/3 of B/D value found in total disk+bulge systems, suggesting bar in disk+bulge systems are preferably found in GOALS galaxies with a large disk and a small bulge.
 
\subsubsection {Strength, size, and ellipticity of bars}

We were able to fit bar components for 19 GOALS nuclei with GALFIT.
The distribution of bar-to-host intensity ratio I$_{bar}$/I$_{host}$ is shown in Fig. 8a. 
The I$_{bar}$/I$_{host}$ ranges from 0.4\% to 30.2\% with mean value of 12.8$\pm$7.8\%.
Similar values are found in nearby field galaxies by Reese et al. (2007, 17.7$\pm$11.3\%) and Weinzirl et al. (2009, 16.6$\pm$10.2\%). Fig. 8b shows the distribution of bar half-light radius r$_e$(bar).
The bar half-light radius in the sample ranges from 0.3 kpc to 6.1 kpc with mean of 2.7$\pm$1.6 kpc, consistent with other studies of bar sizes (e.g., Menendez-Delmestre et al. 2007; Weinzirl et al. 2009).
The similar values of I$_{bar}$/I$_{host}$ and r$_e$(bar) in GOALS galaxies with those found in nearby spirals suggest that bars in galaxies have characteristic ranges of I$_{bar}$/I$_{host}$ and r$_e$(bar) regardless of the source of perturbation.

Fig. 9 shows the distribution of bar ellipticity $e$ (Fig. 9a),
r$_e$ vs. $e$ (Fig. 9b), and B/D vs. $e$ (Fig. 9c) respectively. 
The bar ellipticities in our sample range from 0.36 to 0.88 with an average value of 0.68$\pm$0.14.
The GOALS galaxies value is consistent with what prior studies have found in nearby galaxies surveys: e=0.67$\pm$0.14 (Weinzirl et al 2009),
e=0.50$\pm$0.13 (Menendez-Delmestre et al. 2007), e=0.50$\pm$0.10 (Marinova \& Jogee 2007), and e=0.49$\pm$0.14 (Reese et al. 2007).
It is interesting to note that bar ellipticities derived from GALFIT (this paper \& Weinzirl et al. 2009) are about 30\% 
larger than those derived from other methods (nonparametric decomposition method, Reese et al. 2007;
and ellipse fit method, Menendez-Delmestre et al. 2007; Marinova \& Jogee 2007).  This may reflect the inability of non-decomposition methods to adequately remove the bulge contribution in the measurement of bar ellipticity.


In Fig. 9b, a statistically significant correlation (r=0.61) 
is found between the bar ellipticity and bar half-light radius.
The small bars tend to be round and the large bars tend to be elongated.
The rarity of ellipticities less than 0.4 suggests that either it is difficult to form (or detect) ovals in merging galaxies -- most bars in merging systems are relatively strong.
In Fig. 9c, a weak correlation (r=0.36) is observed between the bar ellipticity and B/D, 
i.e. bars become round (small ellipticity) in galaxies having large bulges.  This is consistent with Das et al. (2003) but we note that this is likely an observational bias because larger bulges lead to apparently fatter bars (e.g., see for example Figure 2 in Sheth et al. 2000).  

In order to find any distinctions that may exist between the barred and unbarred systems,
the mean values of M(host), M(bulge), and B/D have been estimated for both populations (Table 5).
The mean M(host) in barred systems is similar to that in unbarred systems ($\Delta m=0.23$ mag).
On the other hand, the mean M(bulge) in barred systems is about one magnitude fainter ($\Delta m=0.78$ mag) than that in unbarred systems.  This may indicate the stabilizing influence of large bulges in merging galaxies against bar formation (e.g., Mihos \& Hernquist 1996).  However note that the frequency of bars in early type galaxies (more bulge-dominated systems) is higher at higher redshifts (z$\sim$1) than late type galaxies (Sheth et al. 2008) but in the local Universe, the fraction of bars in bright, disk galaxies is roughly the same across Sa-Sc galaxies (Eskridge et al. 2002; Menendez-Delmestre et al. 2007 and references therein; Sheth et al. 2012).  The role of bulges in formation or inhibition of bars during interactions and mergers is worth exploring in more detail with larger samples.


\subsection{Nuclear Point Source}

As galaxies reach more advanced merger stages, more gas will be driven to the central region and 
thereby provide a more favorable environment for compact nuclear starburst and AGN activity.
In this section, the properties of the point source (PSF) component in the GOALS galaxies are examined.
The PSF magnitudes derived from GALFIT are listed in column 8 of Table 2.
The FWHM of the PSF component is about 0.1 arcsec ($\sim$2 pixels) and mean redshift of the GOALS galaxies is z=0.037.
Thus, the maximum physical size of the corresponding region of the PSF emission is $\sim70$ pc, i.e., much smaller than that of the typical compact, nuclear starburst region in (U)LIRGs (a few hundred pc).

\subsubsection {Detection rate of PSF component}

A PSF component has been detected in 22 out of 137 GOALS nuclei (16.1\%).
The fractions of galaxies with a PSF component in GOALS galaxies as a function of interaction class, morphological type, and infrared luminosity (LIRG vs. ULIRG) 
are plotted in Fig. 10 and tabulated in Table 6.
The fraction on top of each histogram represents number of galaxies with PSF detection to the total number of galaxies in the bin.
The PSF fraction in interacting galaxies (IC=1 to IC=4) ranges from 5.9\% to 16.7\%.
When they become more advanced single nucleus mergers, the PSF fraction increases rapidly (33.3\% in IC=5 and 50.0\% in IC=6).
If we compare PSF fractions in interacting galaxies (IC=1 to IC=4) and more advanced single nucleus mergers (IC=5 and IC=6),
we can have more robust statistics in each group with the increased number of data.
The PSF fraction in single nucleus mergers (9/25 = 36.0\%) is about 4 times larger than that in interacting galaxies (10/103 = 9.7\%).
The PSF fraction in isolated systems (IC=0, 33.3\%) is similar to that in single-nucleus mergers (note again that the isolated systems might be post-mergers, but the total number of galaxies in this bin is not sufficient to confirm; only 3 out of 9).

The PSF fractions are 2.3\% (1/44), 22.1\% (15/67), and 25.9\% (6/26) for disk, disk$+$bulge, and elliptical galaxies, respectively.
Compared to the disk galaxies, the PSF detection fractions in disk$+$bulge and elliptical systems are about an order of magnitude larger.
If the PSF component is a compact nuclear starburst, 
the above result suggests that elliptical-like GOALS galaxies have more compact starbursts or unresolved star clusters in the nucleus than in disk-like GOALS galaxies.

The PSF fraction in ULIRGs (25.0\%) is about 1.8 times larger than that in LIRGs (13.8\%).
In general, the molecular hydrogen mass and star formation rate are higher in ULIRGs than in LIRGs.
Large and more concentrated molecular gas in ULIRGs than in LIRGs implies that ULIRGs have more dust extinction than in LIRGs.
This suggests that outflows associated with nuclear activity in ULIRGs is strong enough to initiate clearing out the nuclear gas.

\subsubsection {PSF strength}

Distributions of PSF to host intensity ratios in GOALS galaxies are shown in Fig. 11.
The I$_{PSF}$/I$_{host}$ in GOALS galaxies estimated from I-band images (left panel)
ranges from 0.002 to 0.056 with mean (median) of 0.015$\pm$0.012 (0.014).
The I$_{PSF}$/I$_{host}$ in GOALS elliptical galaxies (cyan line on top-left panel) tend to be larger than the mean value.
We do not see any difference in the I$_{PSF}$/I$_{host}$ between GOALS LIRGs (median=0.015) and GOALS ULIRGs (median=0.013).
The right panel in Fig. 11 shows distributions of I$_{PSF}$/I$_{host}$ estimated from H-band images (Haan et al. 2011) for the GOALS galaxies classified as ellipticals (14 objects listed in Table 8).
We might expect that the I$_{PSF}$/I$_{host}$ would be larger as measured in the H-band compared to that measured in the I-band 
simply due to the reduced extinction in the near-infrared (extinction at H is 1/3 less than at I).  
Among the GOALS LIRGs we see little or no difference between the I-band and H-band data, 
but the GOALS ULIRGs appear to have a larger ratio as measured in the H-band compared to the GOALS LIRGs seen in both filters. 
The median I$_{PSF}$/I$_{host}$ among GOALS ULIRGs is 0.033 while only 0.008 among GOALS LIRGs.

Fig. 12a shows mean I$_{PSF}$/I$_{host}$ as a function of morphological type.
The I$_{PSF}$/I$_{host}$ of GOALS elliptical hosts is about twice as large as that of disk+bulge hosts.
One of the main differences between disk+bulge and elliptical systems is the bulge fraction, and
if the ellipticals have stronger PSF emission than disk+bulge systems, the PSF emission might be correlated with the bulge mass.
Fig. 12b shows absolute bulge magnitude versus absolute PSF magnitude plot. 
We find a marginally good correlation (r=0.54) between PSF and bulge magnitudes.
Correlations have been measured between the black hole mass and bulge magnitude and the nuclear star cluster magnitude and bulge magnitude (Ferrarese et al. 2006); the PSF-bulge magnitude correlation observed for our sample are consistent with both.

\subsection {Tidal features}

Tidal features such as tails, bridges, lopsided disks, distorted outer isophotes, and off-center nuclei
are visible in almost all of the GOALS galaxies.
The strength of the tidal features has been estimated from the residual maps in Fig. 1 and
the intensity ratio of residual to host galaxies is listed in Column 12 of Table 2. 
Distributions of residual to host intensity ratios in GOALS galaxies (red), GOALS LIRGs (green), and GOALS ULIRGs (blue) are shown in Fig. 13.
The I$_{res}$/I$_{host}$ in GOALS galaxies ranges from 0.18 to 0.71 with mean (median) of 0.38$\pm$0.11 (0.36).
We do not find any difference in I$_{res}$/I$_{host}$ between GOALS LIRGs (mean (median)=0.37$\pm$0.11 (0.35)) and 
GOALS ULIRGs (mean (median)=0.40$\pm$0.12 (0.38)).
These numbers are, of course, consistent with GOALS galaxies being greatly disturbed systems.

Fig. 14 shows I$_{res}$/I$_{host}$ of each component in GOALS binary systems as a function of host absolute magnitude.
In this Figure, binary pairs are connected with a solid line.
The circles with the red bar and the green filled circles represent galaxies with a stellar bar and galaxies with a PSF, respectively.
In a binary system, if the I$_{res}$/I$_{host}$ in a small companion (faint component) is larger than that in large one (bright component), it is plotted in Fig. 14a
and the opposite case is plotted in Fig. 14b.
In GOALS binary systems, the fraction of the systems
where the small companion has a larger I$_{res}$/I$_{host}$ (suffering more disturbance) than the large one is 60\%,
and it is 40\% for the opposite case.
About 30\% more bars and about twice as many PSFs are found in the binary systems 
where the small companion suffers more disturbance than the large one (galaxies plotted in Fig. 14a).
The mean I$_{res}$/I$_{host}$ values of the binary systems in Fig. 14a (I$_{res}$/I$_{host}$=0.41$\pm$0.18) 
and in Fig. 14b (I$_{res}$/I$_{host}$=0.36$\pm$0.15) do not show any significant difference.

\section {Discussion}

\subsection {Trend With Nuclear Separation}

In this section, we will investigate trend of morphological parameters as a function of nuclear separation.
For this analysis, we will use single and double nucleus systems (isolated systems, IC=0, are not included).
The nuclear separation is divided into 4 bins in order to have a comparable number of data points (0 kpc, 0-10 kpc, 10-30 kpc, 30-70 kpc).

We have plotted 9 basic morphological parameters as a function nuclear separation in Fig. 15
(fraction of morphological types (a), host magnitude (b), I$_{bulge}$/I$_{host}$ (c), bar fraction (d), I$_{bar}$/I$_{host}$ (e), bar size (f), PSF fraction (g), I$_{PSF}$/I$_{host}$ (h), and I$_{res}$/I$_{host}$ (i)).

Simulations suggest that strong starburst activity commences in interacting galaxies after their first pericentric encounter (which is around 50 kpc, see details in Gerritsen 1997; Mihos \& Hernquist 1996).
At this distance, the dominant morphological type in GOALS galaxies is the disk+bulge system (NS=30-70 kpc in Fig. 15a, disk+bulge systems = $\sim$70\%, and disk galaxies = $\sim$25\%).
If the morphological types of widely separated GOALS galaxies are representative of (U)LIRG progenitors, 
then the majority of progenitors of the GOALS galaxies are disk+bulge systems.
Elliptical hosts represent about 20\% of the galaxies when the nuclear separation of the GOALS galaxies becomes less than 10 kpc. 
The absolute magnitude of the host galaxies as a function of nuclear separation (Fig. 15b) shows no trend. 
The M(host) in galaxies with small nuclear separation could be smaller than that in galaxies with large nuclear separation due to increased extinction. 
However, we do not find any difference of M(host) between galaxies with NS $>$ 10 kpc (M(host)=$-22.51\pm 0.59$) and galaxies with
NS $<$ 10 kpc (M(host)=$-23.07\pm 0.50$).

Although there exist large error bars in I$_{bulge}$/I$_{host}$ vs. NS plot (Fig. 15c), 
we find increasing trend in the median value of I$_{bulge}$/I$_{host}$ with decreasing nuclear separation.
The median values of I$_{bulge}$/I$_{host}$ in each NS bin are 0.18, 0.24, 0.53, and 0.60 
for galaxies with NS=30-70 kpc, NS=10-30 kpc, NS=0-10 kpc, and NS=0 kpc, respectively.
When galaxies are relatively well separated, the I$_{bulge}$/I$_{host}$ increased slightly ($\sim$33\% from NS=50 kpc to NS=20 kpc).
However, when they are close together, a significant increase of I$_{bulge}$/I$_{host}$ occurs (250\% from NS=20 kpc to NS=0 kpc).
If we assume galaxies at NS=20 kpc become completely merged (NS=0 kpc) with radial velocity of 200 km s$^{-1}$,
the I$_{bulge}$/I$_{host}$ doubles in 7.8$\times 10^7$ years.

The bar detection fraction in GOALS galaxies (Fig. 15d) shows a decreasing trend as a function of nuclear separation.
When galaxies are relatively well separated (NS=$30-70$ kpc), about 57\% of the galaxies have a bar.
This fraction is similar to what have been found in other samples of nearby spiral galaxies.
If we consider orientation effects as discussed in $\S 4.3.1$,
the bar fraction of 57\% in well separated galaxies will become 73\% (9/7$\times$57\%), which is about 10\% larger than that in normal galaxies.
Fig. 15d is possible evidence that the bar destruction rate increases with decreasing nuclear separation: 
$\sim 20$\% of bars are destroyed in the first 50 kpc to 20 kpc interval (30 kpc), $\sim 25$\% in the following 15 kpc interval, and $\sim 15$\% in the final 5 kpc interval.
On average, it will take about 1.2$\times10^8$ yrs to destroy half of the bars from NS = 50 kpc NS=0 kpc
if we adopt mean interaction speed of 200 km s$^{-1}$.
This value is about 10\% of the bar decaying time in Sb and Sc galaxies as estimated from numerical simulations
(Bournaud et al. 2005).

If the bars are decaying slowly, we should observe a smooth trend of decreasing I$_{bar}$/I$_{host}$ ratio as a function of nuclear separation among the GOALS galaxies that show bars.
On the other hand, if the bars are decaying abruptly, no dependence of I$_{bar}$/I$_{host}$ with decreasing nuclear separation will be
observed.
As seen in Fig. 15e, we do not see any dependence of I$_{bar}$/I$_{host}$ on the nuclear separation, suggesting
bar destruction could occur rather abruptly.
Similarly, the fact that we see no correlation of the bar half light radius with nuclear separation argues against a slow decay of existing bars (Fig. 15f).

The fraction of GOALS galaxies with PSF component as a function of nuclear separation is
plotted in Fig. 15g and tabulated in Table 6.
The PSF detection fraction for widely separated galaxies (NS = $30-70$ kpc) is about 17\% whereas it is about 8\% for closely interacting (NS=10-30 kpc) and overlapping disk galaxies (NS=0-10 kpc).
The PSF detection fraction is expected to increase with decreasing nuclear separation 
since the gas that fuels vigorous starburst or related AGN activity will be more centrally-concentrated in the nuclear region.
An additional contributing factor that causes the opposite effect is obscuration.
So, compared to the widely separated galaxies, the small PSF detection fraction in closely interacting and overlapping disk galaxies
might be caused by increased extinction.
This can become more apparent if we compare PSF detection fractions of the same objects observed in I and H bands.
For the GOALS elliptical galaxies, we find a PSF detection fraction of 26.3\% with I-band images and 73.7\% with H-band images of Haan et al. (2011).
If $\alpha = 1.5$ is assumed for the extinction law of $A_\lambda \propto \lambda^{-\alpha}$,
extinction at $I$-band is about 3 times larger than that in $H$-band.
We find a sharp rise of the PSF detection fraction from overlapping disk galaxies to single-nucleus merged galaxies (8\% to 38\%, i.e., a $~5$ times increase).
The mean I$_{PSF}$/I$_{host}$  increases from widely-separated galaxies (NS=30-70 kpc) to overlapping disk galaxies (NS=0-10 kpc), 
but decreases at NS=0 kpc probably because of strong dust obscuration in the nuclear region (Fig. 15h).
Note, however, that the large error bars also allow for the possibility that there is no change in 
I$_{PSF}$/I$_{host}$ as a function of nuclear separation.

A low overall fraction of light in tidal features is expected as a function of decreasing nuclear separation.
However, we do not see such trend in I$_{res}$/I$_{host}$ vs. nuclear separation plot (Fig. 15i),
suggesting a significantly long time is required to settling down the tidal features even when they are completely merged.
A similar result was also observed in 1 Jy ULIRGs, where 78.5\% (51 out of 65) of 
completely merged ULIRGs (Veilleux et al. 2002) show prominent tidal features.
Simulations (Hibbard \& Mihos 1995) and deep optical \& radio observations (Hibbard \& van Gorkom 1996) of interacting galaxies 
suggest that the bases of the tails fall back quickly to galaxies, while the more distant regions fall back more slowly.
This delayed return of tidally ejected material may evolve on very long time scales, it may leave observable signatures for many Gyrs (Hibbard \& van Gorkom 1996) as evidenced in tidal features of (U)LIRGs. 
The infalling tidal tails could be the source of disk growth over long period of time (Barnes 2002).

\subsection {Kormendy Relation}

Fig. 16 shows surface brightness vs. half-light radius for the elliptical hosts of the GOALS galaxies (red circles) 
and SDSS ellipticals (green dots; Bernardi et al. 2003 using $I-z$=1.00).
The least-square fit of the GOALS ellipticals (red line) shows similar slope as that of the $z$-band data of Bernardi et al. (green line).
The mean size of GOALS galaxies is r$_{1/2}$=5.3$\pm$3.7 kpc, which is comparable to that of the SDSS ellipticals (r$_{1/2}$=5.2$\pm$3.0 kpc).
These two results may be an indication that the GOALS galaxies may belong to the same parent population as the SDSS ellipticals.

Our first paper (Haan et al. 2011) contains fitting results for 19 GOALS ellipticals presented in this paper.
Haan et al. allow the Sersic index to vary. 
To be consistent with the present I-band analysis, we refit the 19 galaxies with GALFIT using the criteria outlined in Section 3.
The result is presented in Table 8 and r$_{1/2}$ vs. $<\mu_{1/2}>$ plot for these ellipticals in I-band (red circles) and in H-band
(brown squares using $I-H=2.65$) are presented in Fig. 16b where the same object is connected with a dotted line. 

In general, the r$_{1/2}$ estimated from H-band is about 2.4 kpc smaller than that from I-band (medians for r$_{1/2}$ (I) and r$_{1/2}$ (H) are 4.8 kpc and 2.4 kpc respectively),
but no correlation is found between the r$_{1/2}$ (I) and r$_{1/2}$ (H) as shown in Fig. 17a.
In contrast, a weak correlation was found between $<\mu_{1/2}>(I)$ and $<\mu_{1/2}>(H)$ (Fig. 17b, r=0.46).
We also find that the difference in half-light radius at I and H bands has a good correlation with the I-H color of the galaxy;
the difference in the half-light radius become smaller for the objects with red I-H color (Fig. 17c, r=0.75).
This suggests that the objects with blue I-H color tend to have strong nuclear emission at H that in turn causes
a large difference between r$_{1/2}$ (I) and r$_{1/2}$ (H).
As shown in Fig. 17d, a strong correlation exists between the host magnitude of m$_I$ and m$_H$ (r=0.90).
The mean (median) of m$_I$ - m$_H$ in the plot is 1.65$\pm$0.63 (1.78) mag.
This value is slightly larger than that found in SDSS ellipticals ($I-H=1.41$ mag, Jahnke et al. 2004)
probably because the GOALS ellipticals have more dust and are thus redder.

\subsection{Comparison with Prior Work}

In an attempt to study the evolutionary connection between ULIRGs and QSOs (QUEST: Quasar ULIRG Evolution Study)
Veilleux et al. (2006, 2009a) have performed morphological analysis of ULIRGs and PG QSO hosts using HST NICMOS H-band images.
The QUEST ULIRG sample consists of 26 highly nucleated (i.e., the ratio of the H-band luminosity from the inner 4 kpc to the total luminosity is larger than 1/3) ULIRGs drawn from 1 Jy sample of ULIRGs (Kim \& Sanders 1998).
The majority of these galaxies host strong optical and infrared AGNs and
are therefore AGN-biased and not a representative sample of the 1 Jy ULIRGs.

One of the main distinctions of morphological types between the GOALS sample and QUEST sample is that 
elliptical fraction and spiral fraction in QUEST sample are larger than 50\% (74\% and 54\% in QUEST ULIRGs and PG QSO hosts, see Table 3) 
and less than 20\% (16\% and 4\% in ULIRGs and PG QSO hosts) respectively.
In contrast, the elliptical fraction in GOALS (U)LIRGs is only 5\% when they are widely-separated and $\sim$20\% when they become single-nucleus mergers.
Most of the QUEST galaxies are single-nucleus advanced mergers, and if all of the LIRGs evolve to ULIRGs,
a significant fraction of the LIRGs turned into ellipticals only when they are completely merged.

Interestingly, despite the large difference in morphological types between GOALS and QUEST samples,
we do not see any difference in host galaxy magnitudes.
The I-band host absolute magnitude of the QUEST ULIRGs is -22.48$\pm$0.65 
which is similar to that of the GOALS galaxies.
A K-S test shows that the host absolute magnitudes of the GOALS galaxies are
similar (P(null)=0.42) to that of the QUEST ULIRGs.
The distribution of absolute magnitudes of the PG QSO hosts is also not much different from that of the GOALS galaxies.
The host absolute magnitudes for the PG QSOs range from -21.54 to -24.43 mag with mean value of -22.87$\pm$0.78 mag.
A K-S test of the absolute magnitudes between the GOALS galaxies and PG QSO hosts shows
that both samples are not significantly different (P(null)=0.39).
The negligible differences in the host galaxy magnitudes between the GOALS galaxies, ULIRGs, and PG QSOs suggests 
that the GOALS galaxies satisfy one of the necessary conditions to evolve into ULIRGs \& QSO hosts.

The PSF fraction in GOALS single-nucleus advanced mergers is 38\%.
The PSF fraction in QUEST ULIRGs is 100\% probably because they were selected to be nucleated and therefore more likely to have dominant AGN.
The mean (median) I$_{PSF}$/I$_{host}$ of the QUEST ULIRGs and PG QSO hosts are 0.69$\pm1.56$ (0.10) and 3.52$\pm$4.58 (1.88) respectively.
The median of the I$_{PSF}$/I$_{host}$ of the QUEST ULIRGs and PG QSO hosts are about 8 and 140 times larger than that in GOALS galaxies.
This can be caused by distance effects since the GOALS galaxies are much closer than QUEST ULIRGs and PG QSOs.
The I$_{PSF}$/I$_{host}$ as a function of redshift is plotted in Fig. 18 where red circles, blue circles, and blue filled circles represent 
GOALS galaxies, QUEST ULIRGs, and PG QSO hosts respectively.
We find no distance dependency of I$_{PSF}$/I$_{host}$ in GOALS and QUEST ULIRGs, except for a weak correlation observed in PG QSO hosts.
The extinction in $I$-band is about 3 times larger than that in H-band.
Thus, even if extinction is considered, the I$_{PSF}$/I$_{host}$ in GOALS galaxies and QUEST ULIRGs are significantly smaller than that in PG QSO hosts.

Fig. 19 shows I$_{PSF}$/I$_{host}$ as a function of $f_{25}/f_{60}$ (Fig. 19a) and infrared luminosity (Fig. 19b)
where red circles, red circles with blue dot, blue circles, and blue filled circles represent
GOALS LIRGs (I-band), GOALS ULIRGs (I-band), QUEST ULIRGs, and PG QSO hosts, respectively.
The arrows represent either upper or lower limits and
the circles with small, medium, and large crosses
representing ULIRGs with outflow velocities less than
1000 km s$^{-1}$ (small crosses, F00188$-$0856, F03250$+$1606, F04313-1649, F05189-2524, F09039+0503, F09539+0857,
F11506+1331, F14197+0813, F20414-1651; Rupke et al. 2005a),
1000 to 5000 km s$^{-1}$ (medium cross, F05024-1941; Rupke et al. 2005b), and
5,000 to 10,000 km s$^{-1}$ (large crosses, F07599+6508, F12540+5708; Rupke et al. 2005b), respectively.
If we consider QUEST ULIRGs and PG QSO hosts only, we find a strong correlation (solid line in Fig. 19a, r=0.78) between $f_{25}/f_{60}$ and I$_{PSF}$/I$_{host}$. 
The IRAS color ratio of $f_{25}/f_{60}$ is a well known indicator of AGN activity in infrared galaxies (e.g., de Grijp et al. 1985, Miley et al. 1985). 
Thus, this correlation could indicate that the PSF strength is directly related to the AGN activity.
For a given $f_{25}/f_{60}$, the I$_{PSF}$/I$_{host}$ in GOALS galaxies estimated from I-band is about an order of magnitude smaller than that in QUEST ULIRGs.
This can be explained by a combined effect of larger extinction at $I$-band (GOALS galaxies) than at $H$-band (QUEST ULIRGs) and/or
a relatively weak AGN strength in GOALS galaxies than in QUEST ULIRGs.

In Fig. 19b, if only the GOALS galaxies and QUEST ULIRGs are considered,
a weak correlation can be found between the $L_{\rm IR}$ and I$_{PSF}$/I$_{host}$ (r=0.48).
However, if the PG QSO hosts are included,
the correlation disappears.  
The PSF strength in QUEST ULIRGs tend to be proportional to the outflow velocity.
The object with the largest outflow velocity has the strongest PSF strength.
The intense starburst in a ULIRG will inevitably power strong superwinds which in turn
will clear out gas and dust in the nuclear region.
If such an event occurs, the spectral energy distribution in a ULIRG no longer peaks
in the far-IR wavelength region; the peak instead shifts to optical/UV wavelengths. As a result, the ULIRGs will become proto-quasars (Sanders et al. 1988; Veilleux et al. 2006, 2009a, 2009b).

The mean (median) I$_{res}$/I$_{host}$ in GOALS galaxies, QUEST ULIRGs, and PG QSO hosts are 
0.38$\pm$0.11 (0.36), 0.20$\pm0.09$ (0.20), and 0.14$\pm$0.06 (0.13) respectively.
There is a clear tendency for decreasing I$_{res}$/I$_{host}$ from GOALS galaxies to QUEST ULIRGs and QUEST ULIRGs to PG QSO hosts.
The I$_{res}$/I$_{host}$ in GOALS ULIRGs is about twice that of QUEST ULIRGs.
Compared to the $H$-band images, the $I$-band images are more sensitive to the detection of tidal features.
In addition, the GOALS ULIRGs (mean NS=4.0$\pm$9.6 kpc) are less evolved than the QUEST ULIRGs (mean NS=1.0$\pm$4.0 kpc).
The combined effect might explain the I$_{res}$/I$_{host}$ difference between the GOALS ULIRGs and QUEST ULIRGs.
The average I$_{res}$/I$_{host}$ in PG QSO hosts is about 30\% less than that in QUEST ULIRGs.
This indicates that the PG QSO hosts are in general more evolutionary advanced systems than the QUEST ULIRGs.

\section{Summary}

{\it Hubble Space Telescope} ACS/WFC F814W-band imaging data has been used to carry out a 2-dimensional
structural analysis of 85 $L_{\rm IR} \geq 10^{11.4}$ L$_\odot$ (U)LIRGs in GOALS sample.
The main results of the study are as follows:

1. The fraction of GOALS galaxies best fit by disk, disk+bulge, and elliptical profiles is 32.1\%, 48.9\%, and 19.0\%, respectively.
The mean host absolute magnitude of the GOALS galaxies is $-22.64\pm$0.62 mag, which corresponds to 1.8$^{+1.4}_{-0.4}$ L$^*_I$.
We do not find any difference of the mean host absolute magnitudes between the GOALS LIRGs and ULIRGs.

2.  The mean bulge absolute magnitude of GOALS galaxies is about 1.1 magnitude fainter than the mean host magnitude.
The median B/D for the GOALS disk+bulge systems is 0.67 which corresponds to that of the Sa-Sab galaxies.
Mass ratios in the GOALS binary systems suggest that most of the galaxies (93\%) are the result of major mergers ($m_{secondary}/m_{primary} > 1/4$).

3. The average bar fraction in GOALS galaxies (26\%) is about one third of that found in nearby spirals ($\sim$65-70\%).
The bar fraction in widely separated (NS=50 kpc) GOALS galaxies is about 56\%, but may be as high as 72\% if orientation effects are taken into account.
We find twice as many bars in GOALS disk+bulge systems (32.8\%) than in pure-disk galaxies (15.9\%).
But most of the disk+bulge systems that contain bars are disk-dominated,
suggesting that bars in disk+bulge systems are preferably found in large disks with small bulges.
This suggests that large bulges in merging galaxies may be stabilizing the disk against bar formation.  But note that this trend is different from the bar fraction in nearby normal, bright disk galaxies (bar fractions are about the same across Sa-Sc types).  This trend is also opposite of the behavior seen at high redshifts where bars and bulges are coeval at z$\sim$1 -- bars are very scarce in very late type galaxies at high redshifts but evolve the fastest from z$\sim$1 to the present to reach parity with early type systems.

The physical properties of bars in GOALS galaxies such as the bar-to-host intensity ratio (I$_{bar}$/I$_{host}$=12.8$\pm$7.8\%),
bar half-light radius (r$_e$(bar)=2.7$\pm$1.6 kpc), and bar ellipticity (e$_{bar}$=0.68$\pm$0.14)
are similar to those found in nearby spiral galaxies.
We do not find any trend of I$_{bar}$/I$_{host}$ and r$_e$(bar) as a function of nuclear separation,
which may be an indication that bar destruction in merging galaxies occurs rather rapidly.

4. A nuclear PSF component has been detected in 16\% of the GOALS galaxies,
with an average PSF-to-host intensity ratio of I$_{PSF}$/I$_{host}$=0.01. The fraction of detected nuclear PSFs 
increases toward later merger stages and is highest in systems with a nuclear separation of 0 kpc. 
Thus, light from an AGN and/or compact nuclear star cluster is more visible
at I-band or AGN activity is more frequent as (U)LIRGs enter their latter stages of evolution.

5. The GOALS galaxies have significant fraction of their light in tidal features (I$_{res}$/I$_{host}$ = 0.38$\pm$0.11).
Examination of the residual strength in GOALS binary systems suggests
small companions suffer more tidal impact than larger companions (about 50\% more).
We do not observe a rapid decrease in tidal features in GOALS galaxies as a function of decreasing nuclear separation,
suggesting a significantly long time is required to settling down the tidal features even when the progenitor
galaxies are completely merged.
The late-stage mergers whose profiles are well fit by a Sersic index of n=1 may be in the process of reforming disks from infall.

6. The mean size (r$_{1/2}$=5.3$\pm$3.7 kpc) and slope of the surface brightness vs. half-light radius relation (i.e., the Kormendy Relation)
in GOALS elliptical hosts are similar to those in nearby SDSS ellipticals.
The above results are consistent with the possibility that the GOALS galaxies belong to the same parent population as the SDSS ellipticals.

The authors thank the anonymous referee for comments and suggestions that greatly improved this paper. 
We also thank C. Peng and G. Soutchkova for useful discussions and assistance.  
DCK, ASE, GCP and TV were supported by NSF grants AST
02-06262 and 1109475, and by NASA through grants HST-GO10592.01-A and HST-GO11196.01-A
from the SPACE TELESCOPE SCIENCE INSTITUTE, which is operated by the Association of Universities for Research in
Astronomy, Inc., under NASA contract NAS5-26555.
This research has made use of the NASA/IPAC Extragalactic Database
(NED) which is operated by the Jet Propulsion Laboratory, California
Institute of Technology, under contract with the National Aeronautics and
Space Administration.

\clearpage

\begin{deluxetable}{lcrcrr|lcrcrr}
\tabletypesize{\scriptsize}
\tablewidth{0pt}
\tablecaption{Basic Properties of the GOALS Sample}
\tablehead{
\multicolumn{1}{c}{Name} &
\multicolumn{1}{c}{IC} &
\multicolumn{1}{c}{NS} &
\multicolumn{1}{c}{$f_{25}/f_{60}$} &
\multicolumn{1}{c}{D$_L$} &
\multicolumn{1}{c|}{${L_{IR} \over L_\odot}$} &
\multicolumn{1}{c}{Name} &
\multicolumn{1}{c}{IC} &
\multicolumn{1}{c}{NS} &
\multicolumn{1}{c}{$f_{25}/f_{60}$} &
\multicolumn{1}{c}{D$_L$} &
\multicolumn{1}{c}{${L_{IR} \over L_\odot}$} \\
\multicolumn{1}{c}{} &
\multicolumn{1}{c}{} &
\multicolumn{1}{c}{kpc} &
\multicolumn{1}{c}{} &
\multicolumn{1}{c}{mpc} &
\multicolumn{1}{c|}{log} &
\multicolumn{1}{c}{} &
\multicolumn{1}{c}{} &
\multicolumn{1}{c}{kpc} &
\multicolumn{1}{c}{} &
\multicolumn{1}{c}{mpc} &
\multicolumn{1}{c}{log}\\
\multicolumn{1}{c}{(1)} &
\multicolumn{1}{c}{(2)} &
\multicolumn{1}{c}{(3)} &
\multicolumn{1}{c}{(4)} &
\multicolumn{1}{c}{(5)} &
\multicolumn{1}{c|}{(6)} &
\multicolumn{1}{c}{(1)} &
\multicolumn{1}{c}{(2)} &
\multicolumn{1}{c}{(3)} &
\multicolumn{1}{c}{(4)} &
\multicolumn{1}{c}{(5)} &
\multicolumn{1}{c}{(6)}
}
\startdata
NGC 0034          & 5 &     0.0 &    0.09 &    84.1 &   11.49  & UGC 08387         & 5 &     0.0 &    0.08 &   110.0 &   11.73 \\
Arp 256           & 2 &    29.9 &    0.16 &   117.5 &   11.48  & NGC 5256          & 3 &     5.8 &    0.15 &   129.0 &   11.56 \\
MCG +12-02-001    & 3 &     5.3 &    0.14 &    69.8 &   11.50  & Arp 240           & 2 &    39.3 &    0.12 &   108.5 &   11.62 \\
IC 1623           & 3 &     5.4 &    0.16 &    85.5 &   11.71  & UGC 08696         & 5 &     0.0 &    0.10 &   173.0 &   12.21 \\
MCG -03-04-014    & 0 &     0.0 &    0.12 &   144.0 &   11.65  & NGC 5331          & 2 &    17.4 &    0.10 &   155.0 &   11.66 \\
CGCG 436-030      & 2 &    33.0 &    0.14 &   134.0 &   11.69  & IRAS F14348-1447  & 4 &     6.2 &    0.08 &   387.0 &   12.39 \\
IRAS F01364-1042  & 5 &     0.0 &    0.07 &   210.0 &   11.85  & IRAS F14378-3651  & 6 &     0.0 &    0.10 &   315.0 &   12.23 \\
III Zw 035        & 3 &     4.7 &    0.08 &   119.0 &   11.64  & VV 340a           & 1 &    27.9 &    0.06 &   157.0 &   11.74 \\
NGC 0695          & 0 &     0.0 &    0.11 &   139.0 &   11.68  & VV 705            & 4 &     6.9 &    0.16 &   183.0 &   11.92 \\
MCG +05-06-036    & 1 &    32.4 &    0.12 &   145.0 &   11.64  & ESO 099-G004      & 3 &     3.7 &    0.14 &   137.0 &   11.74 \\
UGC 02369         & 2 &    13.5 &    0.23 &   136.0 &   11.67  & IRAS F15250+3608  & 5 &     0.0 &    0.18 &   254.0 &   12.08 \\
IRAS F03359+1523  & 3 &     7.7 &    0.11 &   152.0 &   11.55  & UGC 09913         & 5 &     0.0 &    0.08 &    87.9 &   12.28 \\
ESO 550-IG 025    & 2 &    11.4 &    0.19 &   138.5 &   11.51  & NGC 6090          & 4 &     4.4 &    0.19 &   137.0 &   11.58 \\
NGC 1614          & 5 &     0.0 &    0.23 &    67.8 &   11.65  & IRAS F16164-0746  & 5 &     0.0 &    0.06 &   128.0 &   11.62 \\
ESO 203-IG001     & 4 &     8.1 &    0.08 &   235.0 &   11.86  & ESO 069-IG006     & 2 &    65.1 &    0.10 &   212.0 &   11.98 \\
VII Zw 031        & 0 &     0.0 &    0.11 &   240.0 &   11.99  & IRAS F16399-0937  & 3 &     5.3 &    0.13 &   128.0 &   11.63 \\
IRAS F05189-2524  & 6 &     0.0 &    0.26 &   187.0 &   12.16  & NGC 6240          & 4 &     0.8 &    0.15 &   116.0 &   11.93 \\
MCG +08-11-002    & 6 &     0.0 &    0.08 &    83.7 &   11.46  & IRAS F17132+5313  & 2 &     6.6 &    0.11 &   232.0 &   11.96 \\
IRAS F06076-2139  & 3 &     6.0 &    0.10 &   165.0 &   11.65  & IRAS F17138-1017  & 0 &     0.0 &    0.14 &    84.0 &   11.49 \\
ESO 255-IG007     & 2 &    10.7 &    0.16 &   173.0 &   11.90  & IRAS F17207-0014  & 5 &     0.0 &    0.05 &   198.0 &   12.46 \\
AM 0702-601       & 1 &    56.7 &    0.20 &   141.0 &   11.64  & IRAS 18090+0130   & 2 &    46.1 &    0.10 &   134.0 &   11.65 \\
IRAS 07251-0248   & 5 &     0.0 &    0.10 &   400.0 &   12.39  & IC 4687           & 2 &    31.7 &    0.18 &    81.9 &   11.62 \\
IRAS 08355-4944   & 3 &     1.5 &    0.24 &   118.0 &   11.62  & IRAS F18293-3413  & 1 &     5.6 &    0.11 &    86.0 &   11.88 \\
NGC 2623          & 5 &     0.0 &    0.08 &    84.1 &   11.60  & NGC 6670          & 2 &    17.0 &    0.12 &   129.5 &   11.65 \\
ESO 060-IG 016    & 3 &     8.2 &    0.13 &   210.0 &   11.82  & VV 414            & 1 &    38.0 &    0.19 &   113.0 &   11.49 \\
IRAS F08572+3915  & 4 &     6.0 &    0.24 &   264.0 &   12.16  & ESO 593-IG008     & 2 &     2.4 &    0.08 &   222.0 &   11.93 \\
IRAS 09022-3615   & 5 &     0.0 &    0.10 &   271.0 &   12.31  & IRAS F19297-0406  & 5 &     0.0 &    0.09 &   395.0 &   12.45 \\
IRAS F09111-1007  & 1 &    39.2 &    0.11 &   246.0 &   12.06  & IRAS 19542+1110   & 0 &     0.0 &    0.12 &   295.0 &   12.12 \\
UGC 04881         & 2 &     9.5 &    0.10 &   178.0 &   11.74  & IRAS 20351+2521   & 0 &     0.0 &    0.12 &   151.0 &   11.61 \\
UGC 05101         & 5 &     0.0 &    0.09 &   177.0 &   12.01  & CGCG 448-020      & 2 &     7.9 &    0.18 &   161.0 &   11.94 \\
ESO 374-IG 032    & 4 &     2.7 &    0.12 &   148.5 &   11.73  & ESO 286-IG019     & 5 &     0.0 &    0.15 &   193.0 &   12.06 \\
IRAS F10173+0828  & 0 &     0.0 &    0.10 &   224.0 &   11.86  & IRAS 21101+5810   & 2 &     7.5 &    0.10 &   174.0 &   11.81 \\
NGC 3256          & 5 &     0.0 &    0.15 &    38.9 &   11.64  & ESO 239-IG002     & 5 &     0.0 &    0.16 &   191.0 &   11.84 \\
IRAS F10565+2448  & 2 &    22.9 &    0.10 &   197.0 &   12.08  & IRAS F22491-1808  & 5 &     0.0 &    0.10 &   351.0 &   12.20 \\
MCG +07-23-019    & 2 &    11.0 &    0.11 &   158.0 &   11.62  & NGC 7469          & 2 &    26.3 &    0.22 &    70.8 &   11.65 \\
IRAS F11231+1456  & 1 &    50.8 &    0.10 &   157.0 &   11.64  & ESO 148-IG002     & 4 &     4.2 &    0.15 &   199.0 &   12.06 \\
NGC 3690          & 3 &     5.0 &    0.22 &    50.7 &   11.93  & IC 5298           & 0 &     0.0 &    0.21 &   119.0 &   11.60 \\
IRAS F12112+0305  & 4 &     4.4 &    0.08 &   340.0 &   12.36  & ESO 077-IG014     & 2 &    14.2 &    0.08 &   186.0 &   11.76 \\
IRAS 12116-5615   & 0 &     0.0 &    0.12 &   128.0 &   11.65  & NGC 7674          & 1 &    18.9 &    0.36 &   125.0 &   11.56 \\
CGCG 043-099      & 5 &     0.0 &    0.09 &   175.0 &   11.68  & IRAS F23365+3604  & 5 &     0.0 &    0.13 &   287.0 &   12.20 \\
ESO 507-G070      & 6 &     0.0 &    0.06 &   106.0 &   11.56  & IRAS 23436+5257   & 4 &     3.3 &    0.13 &   149.0 &   11.57 \\
IRAS 13120-5453   & 5 &     0.0 &    0.07 &   144.0 &   12.32  & MRK 0331          & 1 &    43.6 &    0.14 &    79.3 &   11.50 \\
VV 250a           & 2 &    22.1 &    0.17 &   142.0 &   11.81  &                   &   &         &         &         &         \\
\enddata

\tablenotetext{\ } {{\it Col 1:}\ Object name. See Armus et al. (2009) for details}
\tablenotetext{\ } {{\it Col 2:}\ Interaction class (Haan et al. 2011, Evans et al. 2013)}
\tablenotetext{\ } {{\it Col 3:}\ Nuclear separation in kpc unit}
\tablenotetext{\ } {{\it Col 4:}\ IRAS $f_{25}$ to $f_{60}$ flux ratio}
\tablenotetext{\ } {{\it Col 5:}\ Luminosity distance in mpc unit}
\tablenotetext{\ } {{\it Col 6:}\ Infrared luminosity of the object}

\end{deluxetable}

\begin{deluxetable}{lrrrrrrrcrrrrrr}
\tabletypesize{\scriptsize}
\tablewidth{0pc}
\tablecaption{GALFIT Results}
\tablehead{
\multicolumn{1}{l}{Name} &  
\multicolumn{1}{c}{n} & 
\multicolumn{1}{c}{m$_n$} & 
\multicolumn{1}{c}{r$_{1\over2}$} & 
\multicolumn{1}{c}{$\mu_{1\over2}$} & 
\multicolumn{1}{c}{$b\over a$} & 
\multicolumn{1}{c}{m$_{m}$} & 
\multicolumn{1}{c}{m$_{psf}$} &
\multicolumn{1}{c}{B} &
\multicolumn{1}{c}{$I_{bar} \over{I_{host}}$} &
\multicolumn{1}{c}{$I_{psf} \over{I_{host}}$} &
\multicolumn{1}{c}{$I_{res} \over{I_{host}}$} &
\multicolumn{1}{c}{M$_t$} &
\multicolumn{1}{c}{M$_{host}$} \\
\multicolumn{1}{l}{(1)} &
\multicolumn{1}{c}{(2)} &
\multicolumn{1}{c}{(3)} &
\multicolumn{1}{c}{(4)} &
\multicolumn{1}{c}{(5)} &
\multicolumn{1}{c}{(6)} &
\multicolumn{1}{c}{(7)} &
\multicolumn{1}{c}{(8)} &
\multicolumn{1}{c}{(9)} &
\multicolumn{1}{c}{(10)} &
\multicolumn{1}{c}{(11)} &
\multicolumn{1}{c}{(12)} &
\multicolumn{1}{c}{(13)} &
\multicolumn{1}{c}{(14)}
}
\startdata
NGC 0034           &    1  &   12.82 &     2.7 &  18.99 &  0.87 &   12.42 &   18.13 &    N  & \nodata &     0.4 &    35.3 &  -22.39 &  -22.39  \\
                   &    4  &   13.69 &     0.5 &  16.25 &  0.69 & \nodata & \nodata &\nodata& \nodata & \nodata & \nodata & \nodata & \nodata  \\
Arp 256 N          &    1  &   13.70 &    10.3 &  22.11 &  0.35 &   13.46 & \nodata &    Y  &    15.9 & \nodata &    84.0 &  -22.83 &  -22.83  \\
                   &    B  &   15.23 &     2.8 &  20.82 &  0.37 & \nodata & \nodata &\nodata& \nodata & \nodata & \nodata & \nodata & \nodata  \\
Arp 256 S          &    1  &   13.68 &     3.4 &  19.69 &  0.61 &   13.56 & \nodata &    N  & \nodata & \nodata &    56.0 & \nodata & \nodata  \\
                   &    4  &   16.00 &     0.7 &  18.41 &  0.32 & \nodata & \nodata &\nodata& \nodata & \nodata & \nodata & \nodata & \nodata  \\
MCG +12-02-001 S   &    1  &   14.13 &     3.3 &  21.15 &  0.32 &   13.55 &   17.17 &    Y  & \nodata &     2.9 &    45.7 &  -21.31 &  -21.29  \\
                   &    4  &   14.51 &     2.7 &  21.06 &  0.48 & \nodata & \nodata &\nodata& \nodata & \nodata & \nodata & \nodata & \nodata  \\
MCG +12-02-001 N   &    4  &   14.24 &     6.2 &  22.61 &  0.68 & \nodata & \nodata &\nodata& \nodata & \nodata &    55.5 & \nodata & \nodata  \\
IC 1623 W          &    1  &   13.12 &     1.4 &  17.89 &  0.92 &   13.12 & \nodata &    N  & \nodata & \nodata &    42.1 &  -22.71 &  -22.71  \\
IC 1623 E          &    1  &   12.94 &     5.0 &  20.45 &  0.39 & \nodata & \nodata &    N  & \nodata & \nodata &    40.9 & \nodata & \nodata  \\
MCG -03-04-014     &    1  &   13.52 &     3.1 &  18.89 &  0.85 &   13.05 & \nodata &    N  & \nodata & \nodata &    26.8 &  -22.79 &  -22.79  \\
                   &    4  &   14.18 &     3.9 &  20.04 &  0.57 & \nodata & \nodata &\nodata& \nodata & \nodata & \nodata & \nodata & \nodata  \\
CGCG 436-030 E     &    1  &   15.87 &     7.4 &  23.29 &  0.51 &   15.51 & \nodata &    Y  &    18.7 & \nodata &    54.5 &  -22.20 &  -22.19  \\
                   &    4  &   18.10 &     2.9 &  23.50 &  0.31 & \nodata & \nodata &\nodata& \nodata & \nodata & \nodata & \nodata & \nodata  \\
                   &    B  &   17.30 &     2.6 &  22.47 &  0.39 & \nodata & \nodata &\nodata& \nodata & \nodata & \nodata & \nodata & \nodata  \\
CGCG 436-030 W     &    1  &   13.75 &     4.2 &  19.96 &  0.64 &   13.75 &   18.64 &    N  & \nodata &     1.0 &    45.3 & \nodata & \nodata  \\
IRAS F01364-1042   &    1  &   15.17 &     3.5 &  20.03 &  0.55 &   15.17 & \nodata &    N  & \nodata & \nodata &    36.6 &  -21.54 &  -21.54  \\
III Zw 035 N       &    1  &   15.11 &     1.8 &  19.73 &  0.57 &   14.04 & \nodata &    N  & \nodata & \nodata &    18.0 &  -21.47 &  -21.47  \\
                   &    4  &   14.55 &     1.7 &  19.06 &  0.38 & \nodata & \nodata &\nodata& \nodata & \nodata & \nodata & \nodata & \nodata  \\
III Zw 035 S       &    1  &   16.36 &     1.1 &  19.90 &  0.68 & \nodata & \nodata &    N  & \nodata & \nodata &    24.4 & \nodata & \nodata  \\
NGC 0695           &    1  &   12.81 &     4.8 &  19.19 &  0.95 &   12.39 & \nodata &    N  & \nodata & \nodata &    38.7 &  -23.25 &  -23.25  \\
                   &    4  &   13.63 &     3.4 &  19.29 &  0.83 & \nodata & \nodata &\nodata& \nodata & \nodata & \nodata & \nodata & \nodata  \\
MCG +05-06-036 E   &    1  &   13.34 &     5.0 &  19.75 &  0.62 &   12.93 & \nodata &    Y  &    12.5 & \nodata &    22.5 &  -23.48 &  -23.47  \\
                   &    4  &   14.54 &     1.2 &  17.84 &  0.50 & \nodata & \nodata &\nodata& \nodata & \nodata & \nodata & \nodata & \nodata  \\
                   &    B  &   15.58 &     2.8 &  20.70 &  0.25 & \nodata & \nodata &\nodata& \nodata & \nodata & \nodata & \nodata & \nodata  \\
MCG +05-06-036 W   &    1  &   14.56 &     5.5 &  21.17 &  0.61 &   13.58 &   18.84 &    Y  &    23.3 &     0.6 &    25.1 & \nodata & \nodata  \\
                   &    4  &   14.49 &     2.1 &  18.97 &  0.65 & \nodata & \nodata &\nodata& \nodata & \nodata & \nodata & \nodata & \nodata  \\
                   &    B  &   15.58 &     2.8 &  20.70 &  0.25 & \nodata & \nodata &\nodata& \nodata & \nodata & \nodata & \nodata & \nodata  \\
UGC 02369 N        &    1  &   13.72 &     7.4 &  21.10 &  0.34 &   12.85 & \nodata &    N  & \nodata & \nodata &    18.9 &  -23.22 &  -23.21  \\
                   &    4  &   13.50 &     4.6 &  19.83 &  0.78 & \nodata & \nodata &\nodata& \nodata & \nodata & \nodata & \nodata & \nodata  \\
UGC 02369 S        &    1  &   14.12 &     4.5 &  20.41 &  0.75 &   13.72 &   17.58 &    N  & \nodata &     2.6 &    38.7 & \nodata & \nodata  \\
                   &    4  &   14.99 &     3.4 &  20.68 &  0.71 & \nodata & \nodata &\nodata& \nodata & \nodata & \nodata & \nodata & \nodata  \\
IRAS F03359+1523 E &    1  &   15.08 &     2.9 &  20.18 &  0.24 &   15.08 & \nodata &    N  & \nodata & \nodata &    41.3 &  -21.53 &  -21.53  \\
IRAS F03359+1523 W &    4  &   15.61 &     1.2 &  18.76 &  0.87 & \nodata & \nodata &\nodata& \nodata & \nodata &     5.5 & \nodata & \nodata  \\
ESO 550-IG 025 N   &    1  &   13.91 &     4.6 &  20.21 &  0.57 &   13.34 & \nodata &    N  & \nodata & \nodata &    24.9 &  -22.77 &  -22.77  \\
                   &    4  &   14.31 &     3.6 &  20.07 &  0.67 & \nodata & \nodata &\nodata& \nodata & \nodata & \nodata & \nodata & \nodata  \\
ESO 550-IG 025 S   &    1  &   14.90 &     2.8 &  20.13 &  0.34 &   14.66 & \nodata &    N  & \nodata & \nodata &    16.8 & \nodata & \nodata  \\
                   &    4  &   16.44 &     0.9 &  19.29 &  0.44 & \nodata & \nodata &\nodata& \nodata & \nodata & \nodata & \nodata & \nodata  \\
NGC 1614           &    4  &   11.83 &     5.5 &  20.00 &  0.80 &   11.83 &   16.35 &\nodata& \nodata &     1.5 &    55.5 &  -22.36 &  -22.34  \\
ESO 203-IG001 E    &    1  &   16.72 &     1.6 &  19.62 &  0.49 &   15.05 & \nodata &    N  & \nodata & \nodata &    19.2 &  -21.89 &  -21.89  \\
                   &    4  &   15.31 &     9.0 &  22.01 &  0.61 & \nodata & \nodata &\nodata& \nodata & \nodata & \nodata & \nodata & \nodata  \\
ESO 203-IG001 W    &    4  &   17.05 &     2.5 &  20.96 &  0.48 &   17.05 & \nodata &\nodata& \nodata & \nodata &    35.0 & \nodata & \nodata  \\
VII Zw 031         &    1  &   14.25 &     3.4 &  18.78 &  0.92 &   14.10 &   19.65 &    N  & \nodata &     0.5 &    27.3 &  -22.99 &  -22.99  \\
                   &    4  &   16.29 &     1.0 &  18.22 &  0.80 & \nodata & \nodata &\nodata& \nodata & \nodata & \nodata & \nodata & \nodata  \\
IRAS F05189-2524   &    1  &   14.81 &     8.6 &  21.86 &  0.85 &   13.66 &   18.99 &    N  & \nodata &     0.7 &    35.6 &  -22.81 &  -22.80  \\
                   &    4  &   14.12 &     1.0 &  16.45 &  0.93 & \nodata & \nodata &\nodata& \nodata & \nodata & \nodata & \nodata & \nodata  \\
MCG +08-11-002     &    1  &   12.92 &     4.1 &  20.03 &  0.75 &   12.92 & \nodata &    N  & \nodata & \nodata &    30.2 &  -21.90 &  -21.90  \\
IRAS F06076-2139 N &    1  &   14.28 &     5.1 &  20.47 &  0.78 &   13.90 & \nodata &    Y  & \nodata & \nodata &    16.8 &  -22.45 &  -22.45  \\
                   &    4  &   15.22 &     1.9 &  19.25 &  0.54 & \nodata & \nodata &\nodata& \nodata & \nodata & \nodata & \nodata & \nodata  \\
IRAS F06076-2139 S &    1  &   15.82 &     6.3 &  22.45 &  0.70 &   15.19 & \nodata &    Y  &     5.5 & \nodata &    22.1 & \nodata & \nodata  \\
                   &    4  &   16.28 &     1.1 &  19.04 &  0.82 & \nodata & \nodata &\nodata& \nodata & \nodata & \nodata & \nodata & \nodata  \\
                   &    B  &   18.07 &     3.8 &  23.60 &  0.12 & \nodata & \nodata &\nodata& \nodata & \nodata & \nodata & \nodata & \nodata  \\
ESO 255-IG007 NW   &    4  &   13.73 &     4.6 &  19.58 &  0.80 &   13.73 & \nodata &\nodata& \nodata & \nodata &    34.4 &  -23.10 &  -23.10  \\
ESO 255-IG007 N    &    1  &   15.22 &     2.1 &  19.35 &  0.38 &   14.55 & \nodata &    Y  & \nodata & \nodata &    26.6 & \nodata & \nodata  \\
                   &    4  &   15.39 &     1.8 &  19.18 &  0.53 & \nodata & \nodata &\nodata& \nodata & \nodata & \nodata & \nodata & \nodata  \\
ESO 255-IG007 S    &    1  &   15.09 &     4.4 &  20.84 &  0.27 & \nodata & \nodata &    N  & \nodata & \nodata &    32.5 & \nodata & \nodata  \\
AM 0702-601 N      &    1  &   14.21 &     3.4 &  19.84 &  0.90 &   14.09 &   18.22 &    Y  & \nodata &     2.0 &    33.1 &  -22.74 &  -22.73  \\
                   &    4  &   16.50 &     0.3 &  16.56 &  0.64 & \nodata & \nodata &\nodata& \nodata & \nodata & \nodata & \nodata & \nodata  \\
AM 0702-601 S      &    4  &   13.37 &     4.8 &  19.72 &  0.87 & \nodata & \nodata &\nodata& \nodata & \nodata &    36.6 & \nodata & \nodata  \\
IRAS 07251-0248    &    1  &   17.54 &     1.4 &  19.16 &  0.65 &   15.40 & \nodata &    N  & \nodata & \nodata &    41.3 &  -22.72 &  -22.72  \\
                   &    4  &   15.56 &     6.4 &  20.52 &  0.57 & \nodata & \nodata &\nodata& \nodata & \nodata & \nodata & \nodata & \nodata  \\
IRAS 08355-4944 E  &    4  &   14.70 &     4.1 &  21.10 &  0.62 &   14.70 &   17.81 &\nodata& \nodata &     5.6 &    31.3 &  -21.03 &  -20.98  \\
IRAS 08355-4944 W  &    1  &   15.78 &     2.0 &  20.59 &  0.39 & \nodata & \nodata &    N  & \nodata & \nodata &    41.3 & \nodata & \nodata  \\
NGC 2623           &    1  &   13.64 &     2.2 &  19.38 &  0.57 &   12.33 & \nodata &    N  & \nodata & \nodata &    38.4 &  -22.15 &  -22.15  \\
                   &    4  &   12.71 &    13.6 &  22.40 &  0.56 & \nodata & \nodata &\nodata& \nodata & \nodata & \nodata & \nodata & \nodata  \\
ESO 060-IG 016 E   &    1  &   15.36 &     3.6 &  20.31 &  0.34 &   15.36 & \nodata &    N  & \nodata & \nodata &    25.8 &  -22.55 &  -22.55  \\
ESO 060-IG 016 W   &    1  &   14.60 &    10.3 &  21.82 &  0.16 & \nodata & \nodata &    N  & \nodata & \nodata &    36.6 & \nodata & \nodata  \\
IRAS F08572+3915 SE&    1  &   16.18 &     6.7 &  22.02 &  0.36 &   16.08 & \nodata &    Y  &     9.7 & \nodata &    67.9 &  -21.73 &  -21.73  \\
                   &    B  &   18.73 &     0.4 &  18.38 &  0.50 & \nodata & \nodata &\nodata& \nodata & \nodata & \nodata & \nodata & \nodata  \\
IRAS F08572+3915 NW&    1  &   16.39 &     2.6 &  20.14 &  0.72 & \nodata & \nodata &    N  & \nodata & \nodata &    49.2 & \nodata & \nodata  \\
IRAS 09022-3615    &    4  &   14.36 &     6.1 &  19.96 &  0.82 &   14.36 & \nodata &\nodata& \nodata & \nodata &    53.0 &  -22.81 &  -22.80  \\
IRAS F09111-1007 E &    1  &   14.24 &     7.8 &  20.54 &  0.66 &   14.04 & \nodata &    Y  &    17.0 & \nodata &    29.1 &  -23.38 &  -23.38  \\
                   &    B  &   16.68 &     2.7 &  20.72 &  0.31 & \nodata & \nodata &\nodata& \nodata & \nodata & \nodata & \nodata & \nodata  \\
                   &    B  &   16.78 &     0.2 &  15.49 &  0.55 & \nodata & \nodata &\nodata& \nodata & \nodata & \nodata & \nodata & \nodata  \\
IRAS F09111-1007 W &    1  &   15.42 &     4.4 &  20.49 &  0.70 &   14.72 & \nodata &    Y  & \nodata & \nodata &    27.0 & \nodata & \nodata  \\
                   &    4  &   15.53 &     3.2 &  19.91 &  0.50 & \nodata & \nodata &\nodata& \nodata & \nodata & \nodata & \nodata & \nodata  \\
UGC 04881 NE       &    1  &   13.65 &     7.1 &  20.40 &  0.57 &   13.29 & \nodata &    Y  & \nodata & \nodata &    44.5 &  -23.32 &  -23.32  \\
                   &    4  &   14.68 &     6.9 &  21.37 &  0.42 & \nodata & \nodata &\nodata& \nodata & \nodata & \nodata & \nodata & \nodata  \\
UGC 04881 SW       &    1  &   14.96 &     2.5 &  19.40 &  0.56 & \nodata & \nodata &    N  & \nodata & \nodata &    39.8 & \nodata & \nodata  \\
UGC 05101          &    1  &   14.28 &     4.8 &  20.17 &  0.87 &   13.47 &   19.21 &    N  & \nodata &     0.5 &    28.6 &  -22.89 &  -22.88  \\
                   &    4  &   14.16 &     6.2 &  20.60 &  0.38 & \nodata & \nodata &\nodata& \nodata & \nodata & \nodata & \nodata & \nodata  \\
ESO 374-IG 032 W   &    1  &   15.60 &     1.5 &  19.38 &  0.74 &   13.22 & \nodata &    N  & \nodata & \nodata &    42.1 &  -22.49 &  -22.49  \\
ESO 374-IG 032 E   &    4  &   13.35 &     9.3 &  21.08 &  0.71 & \nodata & \nodata &\nodata& \nodata & \nodata & \nodata & \nodata & \nodata  \\
IRAS F10173+0828   &    1  &   16.00 &     4.3 &  21.22 &  0.17 &   15.13 & \nodata &    N  & \nodata & \nodata &    26.3 &  -21.64 &  -21.64  \\
                   &    4  &   15.77 &     3.5 &  20.52 &  0.35 & \nodata & \nodata &\nodata& \nodata & \nodata & \nodata & \nodata & \nodata  \\
NGC 3256           &    1  &   10.79 &     3.1 &  18.89 &  0.69 &   10.70 & \nodata &    N  & \nodata & \nodata &    47.4 &  -22.59 &  -22.59  \\
                   &    4  &   13.46 &     0.3 &  16.81 &  0.67 & \nodata & \nodata &\nodata& \nodata & \nodata & \nodata & \nodata & \nodata  \\
IRAS F10565+2448 W &    4  &   13.76 &     4.8 &  19.47 &  0.78 &   13.76 & \nodata &\nodata& \nodata & \nodata &    33.4 &  -23.00 &  -23.00  \\
IRAS F10565+2448 E &    4  &   15.38 &     1.3 &  18.24 &  0.90 & \nodata & \nodata &\nodata& \nodata & \nodata &    14.6 & \nodata & \nodata  \\
IRAS F10565+2448 S &    4  &   17.13 &     0.9 &  19.08 &  0.74 & \nodata & \nodata &\nodata& \nodata & \nodata &    13.3 & \nodata & \nodata  \\
MCG +07-23-019     &    1  &   13.94 &     4.4 &  19.90 &  0.25 &   13.94 & \nodata &    N  & \nodata & \nodata &    56.5 &  -22.43 &  -22.43  \\
IRAS F11231+1456 W &    1  &   13.67 &     9.0 &  21.17 &  0.34 &   13.31 & \nodata &    Y  &    17.4 & \nodata &    37.3 &  -23.13 &  -23.13  \\
                   &    4  &   15.95 &     1.0 &  18.61 &  0.65 & \nodata & \nodata &\nodata& \nodata & \nodata & \nodata & \nodata & \nodata  \\
                   &    B  &   15.11 &     6.1 &  21.78 &  0.12 & \nodata & \nodata &\nodata& \nodata & \nodata & \nodata & \nodata & \nodata  \\
IRAS F11231+1456 E &    1  &   14.44 &     3.9 &  20.11 &  0.29 & \nodata & \nodata &    Y  & \nodata & \nodata &    37.3 & \nodata & \nodata  \\
NGC 3690 SW        &    1  &   11.86 &     6.9 &  21.14 &  0.48 &   11.32 &   16.99 &    N  & \nodata &     0.7 &    45.3 &  -22.79 &  -22.78  \\
                   &    4  &   12.34 &     2.4 &  19.37 &  0.55 & \nodata & \nodata &\nodata& \nodata & \nodata & \nodata & \nodata & \nodata  \\
NGC 3690 NE        &    1  &   11.95 &     2.6 &  19.10 &  0.83 & \nodata & \nodata &    N  & \nodata & \nodata &    33.4 & \nodata & \nodata  \\
IRAS F12112+0305 N &    1  &   16.11 &     3.3 &  19.91 &  0.64 &   16.11 & \nodata &    N  & \nodata & \nodata &    49.7 &  -22.68 &  -22.68  \\
IRAS F12112+0305 S &    4  &   15.64 &     9.3 &  21.70 &  0.84 & \nodata & \nodata &\nodata& \nodata & \nodata & \nodata & \nodata & \nodata  \\
IRAS 12116-5615    &    4  &   13.77 &     4.0 &  19.93 &  0.94 &   13.77 &   18.02 &\nodata& \nodata &     2.0 &    22.1 &  -21.78 &  -21.75  \\
CGCG 043-099       &    1  &   15.33 &     2.9 &  20.17 &  0.61 &   13.45 & \nodata &    Y  & \nodata & \nodata &    33.4 &  -22.64 &  -22.64  \\
                   &    4  &   13.66 &    11.9 &  21.56 &  0.74 & \nodata & \nodata &\nodata& \nodata & \nodata & \nodata & \nodata & \nodata  \\
ESO 507-G070       &    1  &   13.36 &     8.2 &  21.47 &  0.30 &   12.36 & \nodata &    N  & \nodata & \nodata &    28.6 &  -22.58 &  -22.58  \\
                   &    4  &   12.91 &    18.8 &  22.82 &  0.41 & \nodata & \nodata &\nodata& \nodata & \nodata & \nodata & \nodata & \nodata  \\
IRAS 13120-5453    &    1  &   13.86 &     7.9 &  21.27 &  0.58 &   13.38 &   18.13 &    N  & \nodata &     1.2 &    32.5 &  -22.47 &  -22.46  \\
                   &    4  &   14.51 &     1.2 &  17.89 &  0.87 & \nodata & \nodata &\nodata& \nodata & \nodata & \nodata & \nodata & \nodata  \\
VV 250a E          &    1  &   16.30 &     1.8 &  20.53 &  0.54 &   14.01 & \nodata &    Y  & \nodata & \nodata &    46.6 &  -22.70 &  -22.70  \\
                   &    4  &   14.15 &     3.0 &  19.45 &  0.51 & \nodata & \nodata &\nodata& \nodata & \nodata & \nodata & \nodata & \nodata  \\
VV 250a W          &    1  &   14.71 &     4.1 &  20.70 &  0.35 &   14.14 &   19.63 &    N  & \nodata &     0.5 &    50.1 & \nodata & \nodata  \\
                   &    4  &   15.11 &     5.6 &  21.80 &  0.68 & \nodata & \nodata &\nodata& \nodata & \nodata & \nodata & \nodata & \nodata  \\
UGC 08387          &    1  &   13.50 &     4.4 &  20.17 &  0.47 &   12.95 & \nodata &    N  & \nodata & \nodata &    39.8 &  -22.48 &  -22.48  \\
                   &    4  &   13.94 &    10.1 &  22.43 &  0.62 & \nodata & \nodata &\nodata& \nodata & \nodata & \nodata & \nodata & \nodata  \\
NGC 5256 NE        &    1  &   13.80 &     4.1 &  20.02 &  0.69 &   13.29 & \nodata &    N  & \nodata & \nodata &    19.8 &  -23.37 &  -23.37  \\
                   &    4  &   14.35 &     1.5 &  18.36 &  0.59 & \nodata & \nodata &\nodata& \nodata & \nodata & \nodata & \nodata & \nodata  \\
NGC 5256 SW        &    4  &   12.71 &     8.3 &  20.45 &  0.79 & \nodata & \nodata &\nodata& \nodata & \nodata &    30.8 & \nodata & \nodata  \\
Arp 240 E          &    1  &   12.41 &     9.2 &  20.73 &  0.36 &   12.04 & \nodata &    N  & \nodata & \nodata &    53.5 &  -23.94 &  -23.94  \\
                   &    4  &   13.39 &     6.5 &  20.96 &  0.60 & \nodata & \nodata &\nodata& \nodata & \nodata & \nodata & \nodata & \nodata  \\
Arp 240 W          &    1  &   12.16 &     6.8 &  19.81 &  0.70 &   12.14 & \nodata &    Y  &     0.4 & \nodata &    55.0 &  -23.94 &  -23.94  \\
                   &    4  &   17.15 &     0.2 &  16.87 &  0.94 & \nodata & \nodata &\nodata& \nodata & \nodata & \nodata & \nodata & \nodata  \\
                   &    B  &   17.96 &     0.4 &  19.20 &  0.28 & \nodata & \nodata &\nodata& \nodata & \nodata & \nodata & \nodata & \nodata  \\
UGC 08696          &    1  &   14.04 &     3.6 &  19.34 &  0.54 &   13.22 & \nodata &    N  & \nodata & \nodata &    44.1 &  -23.17 &  -23.17  \\
                   &    4  &   13.90 &    10.0 &  21.44 &  0.35 & \nodata & \nodata &\nodata& \nodata & \nodata & \nodata & \nodata & \nodata  \\
NGC 5331 N         &    1  &   13.43 &     5.5 &  19.89 &  0.50 &   13.23 & \nodata &    Y  &     4.9 & \nodata &    30.5 &  -23.66 &  -23.66  \\
                   &    4  &   15.58 &     0.4 &  16.47 &  0.60 & \nodata & \nodata &\nodata& \nodata & \nodata & \nodata & \nodata & \nodata  \\
                   &    B  &   16.39 &     2.0 &  20.69 &  0.24 & \nodata & \nodata &\nodata& \nodata & \nodata & \nodata & \nodata & \nodata  \\
NGC 5331 S         &    1  &   13.10 &     7.5 &  20.24 &  0.35 & \nodata & \nodata &    N  & \nodata & \nodata &    38.4 & \nodata & \nodata  \\
IRAS F14348-1447 NE&    1  &   16.07 &     5.3 &  20.67 &  0.41 &   16.07 & \nodata &    N  & \nodata & \nodata &    29.1 &  -23.39 &  -23.39  \\
IRAS F14348-1447 SW&    4  &   14.86 &    14.0 &  21.57 &  0.78 & \nodata & \nodata &\nodata& \nodata & \nodata & \nodata & \nodata & \nodata  \\
IRAS F14378-3651   &    1  &   16.16 &     4.2 &  20.63 &  0.67 &   15.51 &   19.88 &    N  & \nodata &     1.7 &    27.5 &  -22.03 &  -22.01  \\
                   &    4  &   16.37 &     1.7 &  18.88 &  0.90 & \nodata & \nodata &\nodata& \nodata & \nodata & \nodata & \nodata & \nodata  \\
VV 340a S          &    1  &   14.01 &     8.2 &  21.31 &  0.80 &   13.24 & \nodata &    N  & \nodata & \nodata &    26.3 &  -23.56 &  -23.56  \\
                   &    4  &   13.98 &     7.5 &  21.10 &  0.82 & \nodata & \nodata &\nodata& \nodata & \nodata & \nodata & \nodata & \nodata  \\
VV 340a N          &    1  &   13.16 &     8.1 &  20.43 &  0.37 & \nodata & \nodata &    N  & \nodata & \nodata &    35.3 & \nodata & \nodata  \\
VV 705 N           &    1  &   14.61 &     3.2 &  19.56 &  0.78 &   14.33 & \nodata &    Y  & \nodata & \nodata &    40.9 &  -22.94 &  -22.94  \\
                   &    4  &   15.95 &     0.7 &  17.62 &  0.67 & \nodata & \nodata &\nodata& \nodata & \nodata & \nodata & \nodata & \nodata  \\
VV 705 S           &    1  &   14.67 &     4.5 &  20.35 &  0.43 & \nodata & \nodata &    N  & \nodata & \nodata &    57.5 & \nodata & \nodata  \\
ESO 099-G004 N     &    1  &   14.08 &     8.6 &  21.78 &  0.51 &   14.08 & \nodata &    N  & \nodata & \nodata &    38.0 &  -22.25 &  -22.25  \\
ESO 099-G004 S     &    1  &   14.42 &     5.3 &  21.06 &  0.35 & \nodata & \nodata &    N  & \nodata & \nodata & \nodata & \nodata & \nodata  \\
IRAS F15250+3608   &    1  &   15.70 &     1.9 &  18.91 &  0.69 &   14.88 &   19.38 &    N  & \nodata &     1.3 &    40.5 &  -22.34 &  -22.33  \\
                   &    4  &   15.57 &     4.5 &  20.63 &  0.81 & \nodata & \nodata &\nodata& \nodata & \nodata & \nodata & \nodata & \nodata  \\
UGC 09913          &    1  &   12.85 &     4.3 &  19.93 &  0.87 &   11.79 & \nodata &    N  & \nodata & \nodata &    33.7 &  -22.75 &  -22.75  \\
                   &    4  &   12.31 &    15.5 &  22.20 &  0.84 & \nodata & \nodata &\nodata& \nodata & \nodata & \nodata & \nodata & \nodata  \\
NGC 6090 W         &    1  &   15.08 &     1.3 &  18.64 &  0.42 &   15.08 & \nodata &    N  & \nodata & \nodata &    81.7 &  -22.81 &  -22.81  \\
NGC 6090 E         &    4  &   13.05 &     4.8 &  19.48 &  0.88 & \nodata & \nodata &\nodata& \nodata & \nodata &    43.2 & \nodata & \nodata  \\
IRAS F16164-0746   &    1  &   13.79 &     4.2 &  20.05 &  0.55 &   13.79 & \nodata &    N  & \nodata & \nodata &    41.7 &  -22.15 &  -22.15  \\
ESO 069-IG006 N    &    1  &   14.07 &     8.4 &  20.83 &  0.46 &   13.81 & \nodata &    Y  &    17.7 & \nodata &    53.5 &  -23.65 &  -23.65  \\
                   &    B  &   15.49 &     2.5 &  19.64 &  0.42 & \nodata & \nodata &\nodata& \nodata & \nodata & \nodata & \nodata & \nodata  \\
ESO 069-IG006 S    &    1  &   15.40 &    10.9 &  22.73 &  0.35 &   14.31 & \nodata &    N  & \nodata & \nodata &    59.2 & \nodata & \nodata  \\
                   &    4  &   14.80 &     2.7 &  19.13 &  0.82 & \nodata & \nodata &\nodata& \nodata & \nodata & \nodata & \nodata & \nodata  \\
IRAS F16399-0937 N &    1  &   14.23 &     4.1 &  20.47 &  0.70 &   14.23 & \nodata &    N  & \nodata & \nodata &    45.7 &  -22.16 &  -22.16  \\
IRAS F16399-0937 S &    4  &   13.67 &     8.2 &  21.41 &  0.80 & \nodata & \nodata &\nodata& \nodata & \nodata &    32.8 & \nodata & \nodata  \\
NGC 6240           &    1  &   11.75 &    11.1 &  20.34 &  0.45 &   11.67 &   18.22 &    N  & \nodata &     0.2 &    39.8 &  -23.79 &  -23.79  \\
                   &    4  &   14.56 &     0.9 &  17.67 &  0.53 & \nodata & \nodata &\nodata& \nodata & \nodata & \nodata & \nodata & \nodata  \\
IRAS F17132+5313 E &    1  &   14.97 &     3.8 &  19.84 &  0.41 &   14.77 & \nodata &    Y  &    13.7 & \nodata &    22.5 &  -22.75 &  -22.75  \\
                   &    B  &   16.70 &     1.6 &  19.69 &  0.29 & \nodata & \nodata &\nodata& \nodata & \nodata & \nodata & \nodata & \nodata  \\
IRAS F17132+5313 W &    1  &   16.63 &     3.7 &  21.42 &  0.61 &   15.11 & \nodata &    N  & \nodata & \nodata &    31.6 & \nodata & \nodata  \\
                   &    4  &   15.42 &     2.3 &  19.23 &  0.83 & \nodata & \nodata &\nodata& \nodata & \nodata & \nodata & \nodata & \nodata  \\
IRAS F17138-1017   &    1  &   13.57 &     3.8 &  20.51 &  0.59 &   13.24 & \nodata &    N  & \nodata & \nodata &    23.1 &  -21.27 &  -21.27  \\
                   &    4  &   14.68 &     5.6 &  22.43 &  0.46 & \nodata & \nodata &\nodata& \nodata & \nodata & \nodata & \nodata & \nodata  \\
IRAS F17207-0014   &    1  &   14.35 &     3.7 &  19.48 &  0.80 &   14.35 & \nodata &    N  & \nodata & \nodata &    38.0 &  -22.43 &  -22.43  \\
IRAS 18090+0130 E  &    1  &   13.93 &     5.4 &  20.66 &  0.48 &   13.74 & \nodata &    N  & \nodata & \nodata &    47.9 &  -22.51 &  -22.51  \\
                   &    4  &   15.75 &     2.0 &  20.31 &  0.38 & \nodata & \nodata &\nodata& \nodata & \nodata & \nodata & \nodata & \nodata  \\
IRAS 18090+0130 W  &    1  &   14.57 &     5.4 &  21.30 &  0.51 &   14.49 & \nodata &    Y  &     3.9 & \nodata &    80.9 & \nodata & \nodata  \\
                   &    4  &   18.41 &     0.5 &  20.05 &  0.31 & \nodata & \nodata &\nodata& \nodata & \nodata & \nodata & \nodata & \nodata  \\
                   &    B  &   17.85 &     2.8 &  23.18 &  0.17 & \nodata & \nodata &\nodata& \nodata & \nodata & \nodata & \nodata & \nodata  \\
IC 4687 N          &    1  &   12.57 &     3.0 &  19.05 &  0.64 &   12.57 & \nodata &    N  & \nodata & \nodata &    42.9 &  -22.95 &  -22.95  \\
IC 4687 W          &    4  &   13.92 &     1.9 &  19.41 &  0.67 & \nodata & \nodata &\nodata& \nodata & \nodata &    34.4 & \nodata & \nodata  \\
IC 4687 S          &    1  &   13.06 &     3.6 &  19.91 &  0.52 &   12.88 & \nodata &    N  & \nodata & \nodata &    25.4 & \nodata & \nodata  \\
                   &    4  &   14.92 &     1.3 &  19.57 &  0.36 & \nodata & \nodata &\nodata& \nodata & \nodata & \nodata & \nodata & \nodata  \\
IRAS F18293-3413 NW&    4  &   14.56 &     0.7 &  17.79 &  0.61 &   14.56 & \nodata &\nodata& \nodata & \nodata &    14.3 &  -22.33 &  -22.33  \\
IRAS F18293-3413 SE&    1  &   12.71 &     2.4 &  18.56 &  0.64 & \nodata & \nodata &    N  & \nodata & \nodata &    38.0 & \nodata & \nodata  \\
NGC 6670 E         &    1  &   16.25 &     2.0 &  20.93 &  0.59 &   16.25 & \nodata &    N  & \nodata & \nodata &    25.1 &  -22.81 &  -22.81  \\
NGC 6670 N         &    1  &   14.60 &     5.2 &  21.30 &  0.55 &   13.50 & \nodata &    N  & \nodata & \nodata &    23.3 & \nodata & \nodata  \\
                   &    4  &   13.99 &     6.9 &  21.33 &  0.20 & \nodata & \nodata &\nodata& \nodata & \nodata & \nodata & \nodata & \nodata  \\
NGC 6670 W         &    1  &   13.80 &     6.5 &  20.99 &  0.18 & \nodata & \nodata &    N  & \nodata & \nodata &    47.0 &  -22.81 &  -22.81  \\
VV 414 NE          &    1  &   13.82 &     5.6 &  20.96 &  0.69 &   13.05 &   17.41 &    N  & \nodata &     1.7 &    54.0 &  -23.37 &  -23.36  \\
                   &    4  &   13.78 &     2.5 &  19.21 &  0.82 & \nodata & \nodata &\nodata& \nodata & \nodata & \nodata & \nodata & \nodata  \\
VV 414 SW          &    1  &   12.99 &     6.6 &  20.51 &  0.91 &   12.51 & \nodata &    Y  &    17.0 & \nodata &    39.8 & \nodata & \nodata  \\
                   &    4  &   14.42 &     1.6 &  18.86 &  0.84 & \nodata & \nodata &\nodata& \nodata & \nodata & \nodata & \nodata & \nodata  \\
                   &    B  &   14.34 &     3.7 &  20.61 &  0.26 & \nodata & \nodata &\nodata& \nodata & \nodata & \nodata & \nodata & \nodata  \\
ESO 593-IG008 W    &    1  &   14.04 &    10.8 &  21.26 &  0.33 &   14.04 & \nodata &    N  & \nodata & \nodata &    20.3 &  -23.19 &  -23.19  \\
ESO 593-IG008 E    &    1  &   14.79 &     7.1 &  21.09 &  0.20 & \nodata & \nodata &    N  & \nodata & \nodata &    98.2 & \nodata & \nodata  \\
IRAS F19297-0406   &    4  &   15.38 &    10.4 &  21.42 &  0.69 &   15.38 &   19.67 &\nodata& \nodata &     1.8 &    41.3 &  -22.66 &  -22.64  \\
IRAS 19542+1110    &    4  &   15.00 &     1.3 &  17.09 &  0.92 &   15.00 &   19.59 &\nodata& \nodata &     1.5 &    21.7 &  -22.34 &  -22.32  \\
IRAS 20351+2521    &    1  &   13.14 &     7.6 &  20.36 &  0.94 &   12.85 & \nodata &    Y  &    23.3 & \nodata &    34.4 &  -23.04 &  -23.04  \\
                   &    B  &   14.77 &     5.1 &  21.11 &  0.31 & \nodata & \nodata &\nodata& \nodata & \nodata & \nodata & \nodata & \nodata  \\
                   &    B  &   15.89 &     0.6 &  17.62 &  0.61 & \nodata & \nodata &\nodata& \nodata & \nodata & \nodata & \nodata & \nodata  \\
CGCG 448-020 W     &    1  &   13.86 &     5.0 &  20.06 &  0.59 &   13.86 & \nodata &    N  & \nodata & \nodata &    48.8 &  -22.79 &  -22.79  \\
CGCG 448-020 S     &    1  &   14.82 &     1.2 &  17.84 &  1.00 & \nodata & \nodata &    N  & \nodata & \nodata &    23.3 & \nodata & \nodata  \\
CGCG 448-020 E     &    1  &   16.16 &     3.3 &  21.45 &  0.79 & \nodata & \nodata &    N  & \nodata & \nodata &    25.8 & \nodata & \nodata  \\
ESO 286-IG019      &    4  &   13.34 &     7.8 &  20.13 &  0.78 &   13.34 & \nodata &\nodata& \nodata & \nodata &    61.4 &  -22.97 &  -22.97  \\
IRAS 21101+5810 NW &    4  &   14.57 &     9.1 &  21.89 &  0.67 &   14.57 & \nodata &\nodata& \nodata & \nodata &    43.6 &  -21.65 &  -21.65  \\
IRAS 21101+5810 SE &    4  &   16.59 &     1.0 &  19.06 &  0.76 & \nodata & \nodata &\nodata& \nodata & \nodata &     7.7 & \nodata & \nodata  \\
ESO 239-IG002      &    4  &   12.71 &    12.2 &  20.49 &  0.83 &   12.71 &   17.30 &\nodata& \nodata &     1.7 &    34.0 &  -23.55 &  -23.53  \\
IRAS F22491-1808   &    1  &   16.26 &     4.5 &  20.68 &  0.28 &   15.20 & \nodata &    N  & \nodata & \nodata &    62.5 &  -22.78 &  -22.78  \\
                   &    4  &   15.72 &     4.9 &  20.34 &  0.44 & \nodata & \nodata &\nodata& \nodata & \nodata & \nodata & \nodata & \nodata  \\
NGC 7469 S         &    1  &   12.46 &     5.6 &  20.60 &  0.53 &   11.59 & \nodata &    Y  & \nodata & \nodata &    36.3 &  -23.13 &  -23.13  \\
                   &    4  &   12.24 &     1.1 &  16.88 &  0.80 & \nodata & \nodata &\nodata& \nodata & \nodata & \nodata & \nodata & \nodata  \\
NGC 7469 N         &    1  &   12.89 &     4.8 &  20.70 &  0.38 & \nodata & \nodata &    N  & \nodata & \nodata &    54.5 & \nodata & \nodata  \\
ESO 148-IG002 N    &    1  &   14.65 &     2.5 &  18.90 &  0.65 &   13.94 & \nodata &    N  & \nodata & \nodata &    49.2 &  -22.97 &  -22.97  \\
ESO 148-IG002 S    &    4  &   14.73 &     3.8 &  19.90 &  0.75 & \nodata & \nodata &\nodata& \nodata & \nodata & \nodata & \nodata & \nodata  \\
IC 5298            &    1  &   14.33 &     3.6 &  20.43 &  0.34 &   12.79 & \nodata &    Y  &     3.8 & \nodata &    18.5 &  -22.60 &  -22.55  \\
                   &    4  &   13.15 &     4.6 &  19.77 &  0.88 & \nodata & \nodata &\nodata& \nodata & \nodata & \nodata & \nodata & \nodata  \\
                   &    B  &   16.39 &     0.3 &  17.09 &  0.51 & \nodata & \nodata &\nodata& \nodata & \nodata & \nodata & \nodata & \nodata  \\
ESO 077-IG014 N    &    1  &   14.51 &     4.8 &  20.31 &  0.36 &   14.07 & \nodata &    Y  &    15.2 & \nodata &    32.2 &  -23.02 &  -23.02  \\
                   &    4  &   16.16 &     2.0 &  20.05 &  0.84 & \nodata & \nodata &\nodata& \nodata & \nodata & \nodata & \nodata & \nodata  \\
                   &    B  &   15.89 &     1.4 &  18.98 &  0.64 & \nodata & \nodata &\nodata& \nodata & \nodata & \nodata & \nodata & \nodata  \\
ESO 077-IG014 S    &    1  &   15.01 &     3.9 &  20.38 &  0.46 &   14.59 & \nodata &    N  & \nodata & \nodata &    42.9 & \nodata & \nodata  \\
                   &    4  &   15.83 &     4.4 &  21.44 &  0.42 & \nodata & \nodata &\nodata& \nodata & \nodata & \nodata & \nodata & \nodata  \\
NGC 7674 W         &    1  &   12.43 &     8.1 &  20.17 &  0.92 &   12.23 & \nodata &    Y  &     7.0 & \nodata &    30.5 &  -23.50 &  -23.50  \\
                   &    4  &   14.70 &     0.4 &  16.10 &  0.75 & \nodata & \nodata &\nodata& \nodata & \nodata & \nodata & \nodata & \nodata  \\
                   &    B  &   15.12 &     4.2 &  21.43 &  0.25 & \nodata & \nodata &\nodata& \nodata & \nodata & \nodata & \nodata & \nodata  \\
NGC 7674 E         &    1  &   14.87 &     2.5 &  20.06 &  0.59 &   13.89 & \nodata &    N  & \nodata & \nodata &    31.6 & \nodata & \nodata  \\
                   &    4  &   14.45 &     1.1 &  17.94 &  0.65 & \nodata & \nodata &\nodata& \nodata & \nodata & \nodata & \nodata & \nodata  \\
IRAS F23365+3604   &    1  &   14.59 &     5.7 &  19.92 &  0.70 &   14.51 & \nodata &    N  & \nodata & \nodata &    38.4 &  -22.99 &  -22.99  \\
                   &    4  &   17.33 &     0.9 &  18.53 &  0.62 & \nodata & \nodata &\nodata& \nodata & \nodata & \nodata & \nodata & \nodata  \\
IRAS 23436+5257 N  &    1  &   14.24 &     3.7 &  19.91 &  0.47 &   14.07 & \nodata &    N  & \nodata & \nodata &    52.0 &  -22.34 &  -22.34  \\
                   &    4  &   16.19 &     0.4 &  16.96 &  0.89 & \nodata & \nodata &\nodata& \nodata & \nodata & \nodata & \nodata & \nodata  \\
IRAS 23436+5257 S  &    4  &   15.19 &     1.9 &  19.47 &  0.67 & \nodata & \nodata &\nodata& \nodata & \nodata &    18.9 & \nodata & \nodata  \\
MRK 0331 E         &    1  &   14.44 &     3.4 &  21.25 &  0.36 &   12.43 & \nodata &    N  & \nodata & \nodata &    29.1 &  -22.25 &  -22.25  \\
                   &    4  &   12.62 &     1.8 &  18.03 &  0.84 & \nodata & \nodata &\nodata& \nodata & \nodata & \nodata & \nodata & \nodata  \\
MRK 0331 W         &    1  &   14.49 &     4.5 &  21.91 &  0.26 & \nodata & \nodata &    N  & \nodata & \nodata &    36.0 & \nodata & \nodata  \\
\enddata

\tablenotetext{\ } {{\it Col 1:}\ Object name}
\tablenotetext{\ } {{\it Col 2:}\ Best fit model components in each nucleus (1: Sersic index n=1 (disk), 4: Sersic index of n=4 (bulge), B: bar component)}
\tablenotetext{\ } {{\it Col 3:}\ Apparent I-band magnitude for each component}
\tablenotetext{\ } {{\it Col 4:}\ Half-light radius in kpc for each component}
\tablenotetext{\ } {{\it Col 5:}\ Mean surface brightness within half-light radius in I mag arcsec$^{-1}$}
\tablenotetext{\ } {{\it Col 6:}\ Axis ratio}
\tablenotetext{\ } {{\it Col 7:}\ Apparent I-band magnitude for all model components}
\tablenotetext{\ } {{\it Col 8:}\ Apparent I-band magnitude for PSF component}
\tablenotetext{\ } {{\it Col 9:}\ Visually-identified bar (Y). If no bar is visually identified, then a 'N' is placed in this column. If GALFIT can decompose bar, it is indicated as B in column (2)}
\tablenotetext{\ } {{\it Col 10:}\ Bar to host intensity ratio (\%)}
\tablenotetext{\ } {{\it Col 11:}\ PSF to host intensity ratio (\%)}
\tablenotetext{\ } {{\it Col 12:}\ Residual (total - model) to host intensity ratio (\%)}
\tablenotetext{\ } {{\it Col 13:}\ Total absolute magnitude of the system}
\tablenotetext{\ } {{\it Col 14:}\ Host absolute magnitude (total - PSF) of the system}
\end{deluxetable}

\clearpage

\begin{deluxetable}{lccccccc}
\tablecolumns{7}
\tablewidth{0pc}
\tablecaption{Distribution of Morphological Type}
\tablehead{
\multicolumn{1}{l}{Morphological Type} &  \multicolumn{1}{c}{GOALS} & \multicolumn{1}{c}{GOALS LIRG} & \multicolumn{1}{c}{GOALS ULIRG} & \multicolumn{1}{c}{QUEST ULIRG$^1$} & \multicolumn{1}{c}{PG QSO host}}
\startdata
Disk (\%)       & 32.1  & 34.5 & 22.2 & 15.8 & 3.8\\
Disk+Bulge (\%) & 48.9  & 50.0 & 44.4 & 10.5 & 42.3 \\
Elliptical (\%) & 19.0  & 15.5 & 33.3 & 73.7 & 53.8 \\
\enddata
\tablenotetext{\ } {$^1$: These ULIRGs were selected based on having a high degree of nucleation at H-band}
\end{deluxetable}

\begin{deluxetable}{lcccccccc}
\tablecolumns{8}
\tablewidth{0pc}
\tablecaption{Bar fraction}
\tablehead{
\multicolumn{1}{l}{Category} &  \multicolumn{7}{c}{Bar Fraction (\%)} }
\startdata
N.S. (kpc) &  \multicolumn{2}{c}{0}   &  \multicolumn{2}{c}{0 - 10}  &  \multicolumn{2}{c}{10 - 30} &  \multicolumn{2}{c}{30 - 70} \\
              &  \multicolumn{2}{c}{5.0} &  \multicolumn{2}{c}{17.9} &  \multicolumn{2}{c}{41.7} &  \multicolumn{2}{c}{57.1}  \\
\tableline
Int. class & 0    & 1    & 2   & 3    & 4   & 5    & 6    \\
           & 28.6 & 56.3 & 30.8 & 17.7 & 16.7 & 6.3 & 0.0 \\
\tableline
Mor. type & \multicolumn{3}{c}{Disk} & \multicolumn{3}{c}{Disk+Bulge} & Total\\
 & \multicolumn{3}{c}{15.9} & \multicolumn{3}{c}{32.8} & 25.9\\
\enddata
\end{deluxetable}

\begin{deluxetable}{lcccccc}
\tablecolumns{6}
\tablewidth{0pc}
\tablecaption{Statistics of Barred and Unbarred Systems}
\tablehead{
\multicolumn{1}{l}{quantity} & \multicolumn{1}{c}{M(host) mag} & \multicolumn{1}{c}{M(bulge) mag} & \multicolumn{1}{c}{B/D (median)} & \multicolumn{1}{c}{D(\%)} & \multicolumn{1}{c}{D+B(\%)} }
\startdata
barred     &-22.26$\pm$0.77 & -20.45$\pm$1.43 &0.36 & 15.9 & 32.8 \\
unbarred  &-22.03$\pm$0.80 & -21.23$\pm$0.88 &0.87 & 84.1 & 67.2 \\
\enddata
\end{deluxetable}

\begin{deluxetable}{lcccccccc}
\tablecolumns{8}
\tablewidth{0pc}
\tablecaption{PSF Detection Fraction}
\tablehead{
\multicolumn{1}{l}{Category} &  \multicolumn{7}{c}{PSF Detection Fraction (\%)} }
\startdata
N.S. (kpc) &  \multicolumn{2}{c}{0}   &  \multicolumn{2}{c}{0 - 10}  &  \multicolumn{2}{c}{10 - 30} &  \multicolumn{2}{c}{30 - 70} \\
              &  \multicolumn{2}{c}{38.5} &  \multicolumn{2}{c}{7.7} &  \multicolumn{2}{c}{7.1} &  \multicolumn{2}{c}{17.4}  \\
\tableline
Int. class & 0    & 1    & 2   & 3    & 4   & 5    & 6    \\
           & 33.3 & 16.7 & 6.5 & 13.6 & 5.9 & 33.3 & 50.0 \\
\tableline
Mor. type & \multicolumn{2}{c}{Disk} & \multicolumn{2}{c}{Disk+Bulge} & \multicolumn{2}{c}{Elliptical} &Total\\
 & \multicolumn{2}{c}{2.3} & \multicolumn{2}{c}{22.4} & \multicolumn{2}{c}{23.1} &16.1 \\
\tableline
Infrared lum. & \multicolumn{3}{c}{LIRG} & \multicolumn{3}{c}{ULIRG} &\\
 & \multicolumn{3}{c}{13.8} & \multicolumn{3}{c}{25.0} & \\
\enddata
\end{deluxetable}

\begin{deluxetable}{lccccc}
\tablecolumns{5}
\tablewidth{0pc}
\tablecaption{Statistics of PSF and Non-PSF Systems}
\tablehead{
\multicolumn{1}{l}{quantity} & \multicolumn{1}{c}{M(host) mag} & \multicolumn{1}{c}{M(bulge) mag} & \multicolumn{1}{c}{B/D} & \multicolumn{1}{c}{$f_{25}/f_{60}$}}
\startdata
PSF     &-22.30$\pm$0.71 & -21.34$\pm$1.00 & 0.73$\pm$0.48 &  0.16$\pm$0.06  \\
non-PSF &-21.95$\pm$0.89 & -21.08$\pm$1.26 & 1.37$\pm$1.83 &  0.13$\pm$0.05  \\
\enddata
\end{deluxetable}

\begin{deluxetable}{lccrcccrc}
\tabletypesize{\footnotesize}
\tablecolumns{9}
\tablewidth{0pc}
\tablecaption{I and H Bands r$_{1/2}$ and $<\mu_{1/2}>$ for GOALS Galaxies with Eliptical Profiles}
\tablehead{
\multicolumn{1}{l}{} &
\multicolumn{4}{c}{I} &
\multicolumn{4}{c}{H} \\
\cline{2-9} 
\multicolumn{1}{l}{Name} &
\multicolumn{1}{c}{m$_h$} &
\multicolumn{1}{c}{m$_{PSF}$} &
\multicolumn{1}{c}{r$_{1/2}$} &
\multicolumn{1}{c}{$<\mu_{1/2}>$} &
\multicolumn{1}{c}{m$_h$} &
\multicolumn{1}{c}{m$_{PSF}$} &
\multicolumn{1}{c}{r$_{1/2}$} &
\multicolumn{1}{c}{$<\mu_{1/2}>$} \\
\multicolumn{1}{l}{(1)} &
\multicolumn{1}{c}{(2)} &
\multicolumn{1}{c}{(3)} &
\multicolumn{1}{c}{(4)} &
\multicolumn{1}{c}{(5)} &
\multicolumn{1}{c}{(6)} &
\multicolumn{1}{c}{(7)} &
\multicolumn{1}{c}{(8)} &
\multicolumn{1}{c}{(9)}  
}
\startdata
IRAS F03359+1523  & 15.61 & \nodata &  1.2 &  18.76 &  13.65 &  17.77 &   1.4 &  22.79 \\ 
NGC 1614          & 11.83 & 16.35   &  5.5 &  20.00 &  10.90 &  13.44 &   0.2 &  17.94 \\ 
ESO 203-IG001     & 17.05 & \nodata &  2.5 &  20.96 &  15.35 & \nodata&   3.4 &  25.51 \\ 
ESO 255-IG007     & 13.73 & \nodata &  4.6 &  19.58 &  11.62 & \nodata&   2.8 &  21.97 \\ 
AM 0702-601       & 13.37 & \nodata &  4.8 &  19.72 &  11.21 &  18.68 &   4.1 &  22.82 \\ 
IRAS F10565+2448  & 13.76 & \nodata &  4.8 &  19.47 &  12.31 &  16.50 &   1.5 &  21.10 \\ 
IRAS F12112+0305  & 15.64 & \nodata &  9.3 &  21.70 &  15.16 &  16.89 &   1.3 &  22.55 \\ 
IRAS 12116-5615   & 13.77 & 18.02   &  4.0 &  19.93 &  10.93 &  15.46 &   2.7 &  21.82 \\ 
NGC 5256          & 12.71 & \nodata &  8.3 &  20.45 &  11.48 & \nodata&   1.6 &  23.65 \\ 
IRAS F14348-1447  & 14.86 & \nodata & 14.0 &  21.57 &  14.17 &  18.99 &   2.1 &  22.39 \\ 
NGC 6090          & 13.05 & \nodata &  4.8 &  19.48 &  11.39 &  19.00 &   5.4 &  23.63 \\ 
IRAS F16399-0937  & 13.67 & \nodata &  8.2 &  21.41 &  11.89 &  18.08 &   2.8 &  22.88 \\ 
IRAS F19297-0406  & 15.38 & 19.67   & 10.4 &  21.42 &  13.31 &  17.03 &   5.5 &  23.56 \\ 
IRAS 19542+1110   & 15.00 & 19.59   &  1.3 &  17.09 &  13.02 &  16.03 &   0.6 &  19.15 \\ 
ESO 286-IG019     & 13.34 & \nodata &  7.8 &  20.13 &  12.08 & \nodata&   2.8 &  22.24 \\ 
IRAS 21101+5810NW & 14.57 & \nodata &  9.1 &  21.89 &  12.54 &  18.58 &   3.9 &  23.62 \\ 
IRAS 21101+5810SE & 16.59 & \nodata &  1.0 &  19.06 &  14.21 & \nodata&   1.1 &  22.48 \\ 
ESO 239-IG002     & 12.71 & 17.30   & 12.2 &  20.49 &  12.01 &  14.64 &   2.4 &  21.86 \\ 
IRAS 23436+5257   & 15.19 & \nodata &  1.9 &  19.47 &  13.32 &  18.51 &   1.0 &  21.81 \\ 
\enddata

\tablenotetext{\ } {{\it Col 1:}\ Object name}
\tablenotetext{\ } {{\it Col 2:}\ Apparent I-band magnitude for host galaxy}
\tablenotetext{\ } {{\it Col 3:}\ Apparent I-band magnitude for PSF component}
\tablenotetext{\ } {{\it Col 4:}\ I-band half-light radius in kpc}
\tablenotetext{\ } {{\it Col 5:}\ Mean surface brightness within half-light radius in I mag arcsec$^{-1}$}
\tablenotetext{\ } {{\it Col 6:}\ Apparent H-band magnitude for host galaxy}
\tablenotetext{\ } {{\it Col 7:}\ Apparent H-band magnitude for PSF component}
\tablenotetext{\ } {{\it Col 8:}\ H-band half-light radius in kpc}
\tablenotetext{\ } {{\it Col 9:}\ Mean surface brightness within half-light radius in H mag arcsec$^{-1}$}

\end{deluxetable}

\begin{figure*}[ht]
\includegraphics[width=1.0\textwidth]{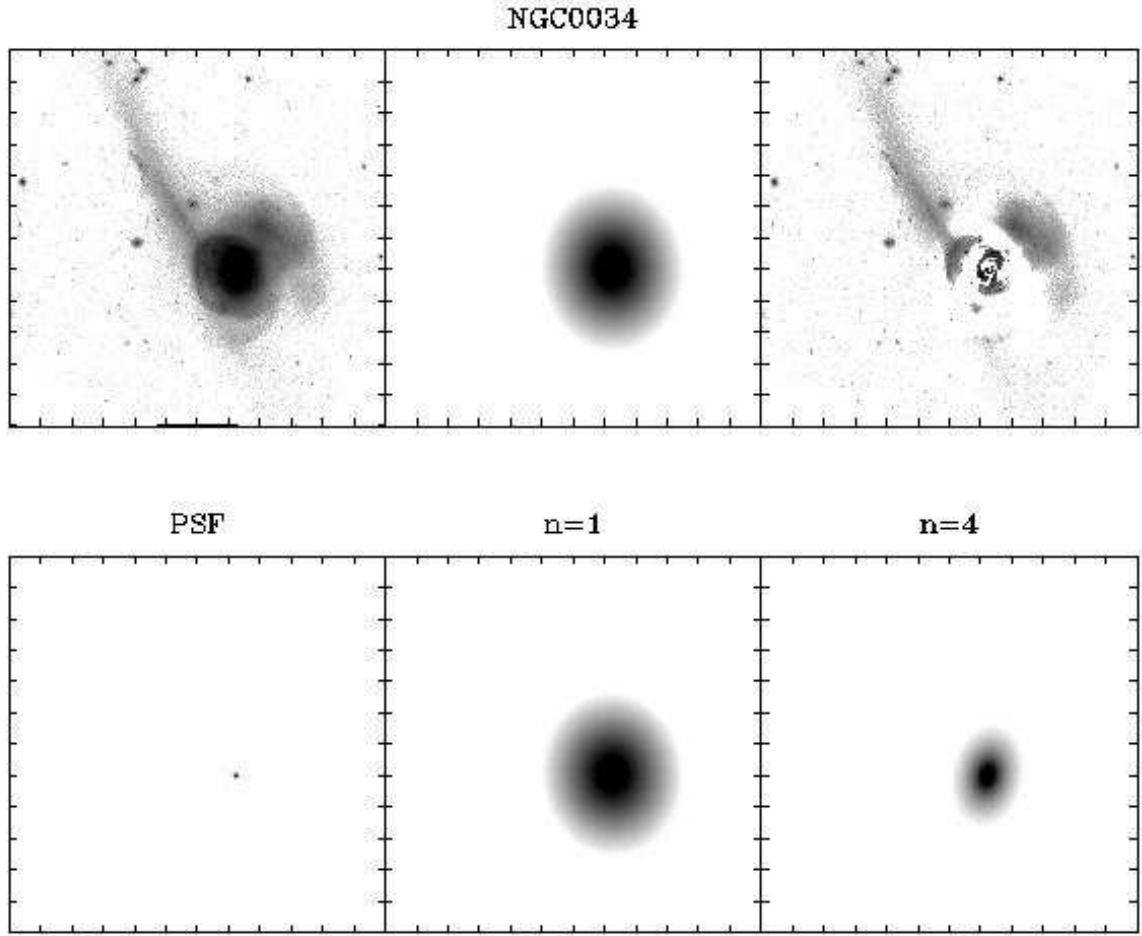}
\caption{Results from the GALFIT analysis.  The upper panel shows the original image (a), model galaxy (b), and residual.
The lower panel shows Sersic components used in the model. 
The component names listed on top of each plot are PSF, disk (n=1), and bulge (n=4). 
The intensity scale is logarithmic and stretch values are the same in all images.
The tick marks in the panels represent 5", and the thick bar on the top-left panel represents 10 kpc scale.
North is top and east is to the left. [\it{See the electronic edition of the Journal for} Figs. 1.2-1.85.]}
\end{figure*}

\begin{figure*}[ht]
\includegraphics[width=1.0\textwidth]{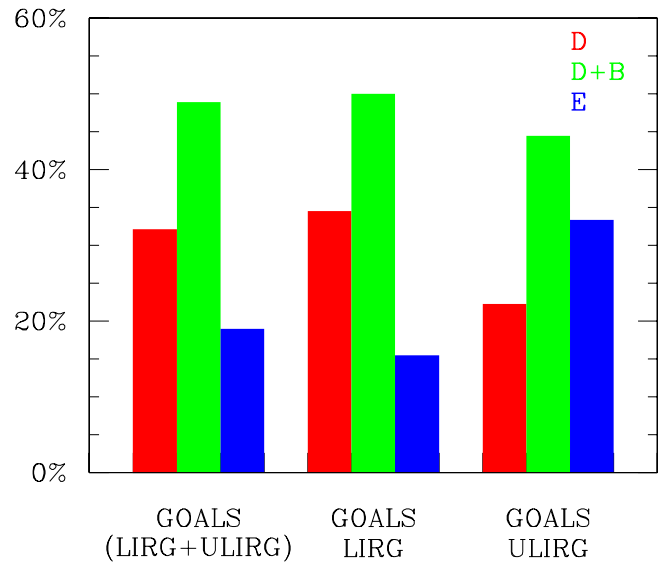}
\caption{ a) Distribution of morphological types in GOALS galaxies, GOALS LIRGs, and GOALS ULIRGs.
The red, green, and blue colors represent disk, disk$+$bulge, and elliptical host respectively. 
}
\end{figure*}

\begin{figure*}
\includegraphics[width=1.0\textwidth]{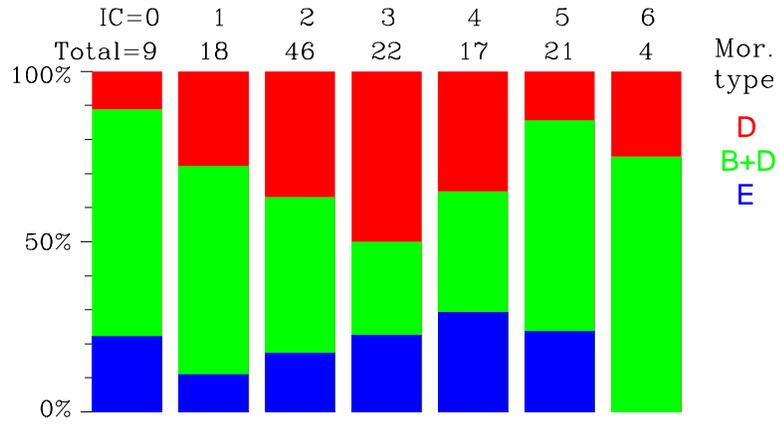}
\caption{ Distribution of morphological types as a function of interaction class.}
\end{figure*}
\clearpage

\begin{figure*}[ht]
\epsscale{1.0}
\includegraphics[width=1.0\textwidth]{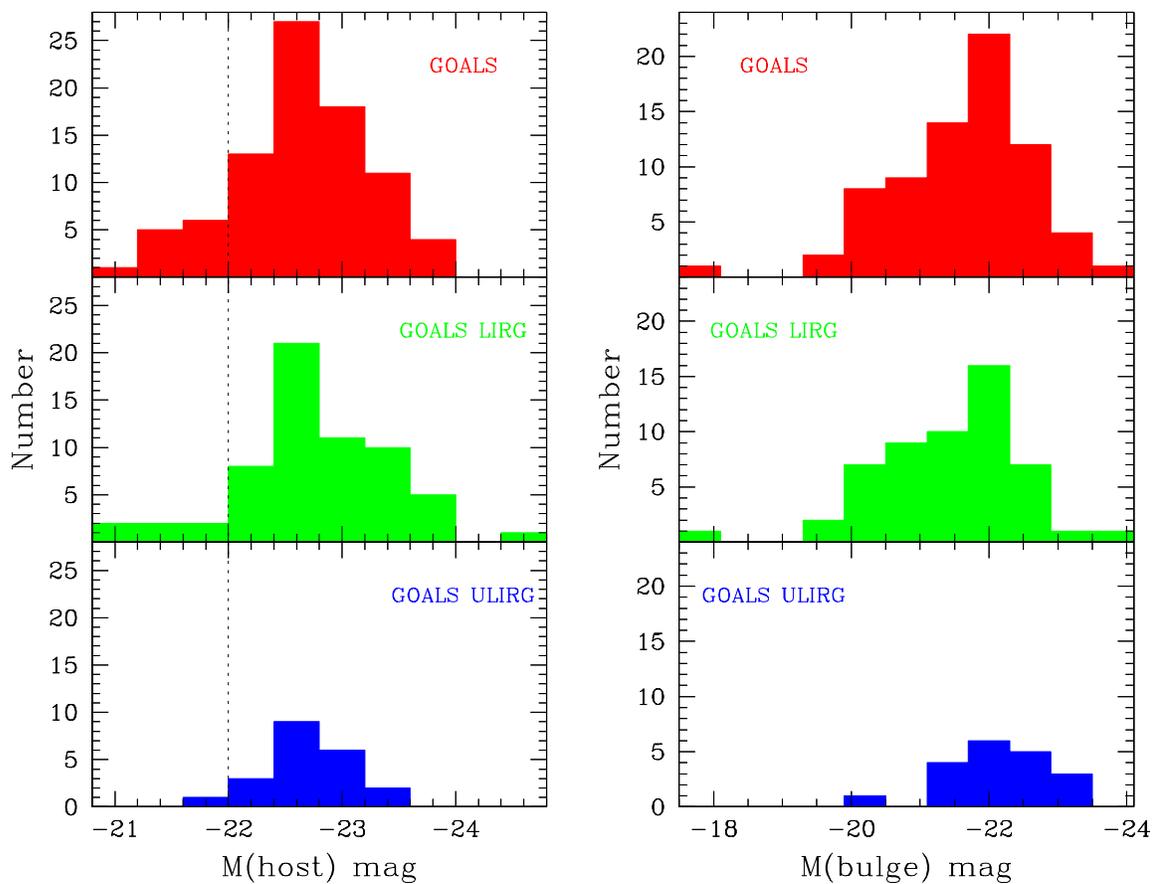}
\caption{ Distribution of a) host absolute magnitude and b) bulge absolute magnitude for GOALS galaxies (red),
GOALS LIRGs (green), and GOALS ULIRGs (blue).
The vertical dashed line represents $M^*_I=-22.0$ mag, i.e., the $I$-band absolute magnitude of an
$L^*$ galaxy in a Schechter luminosity function of the local field galaxies. }
\end{figure*}
\clearpage

\begin{figure*}[ht]
\epsscale{1.0}
\includegraphics[width=1.0\textwidth]{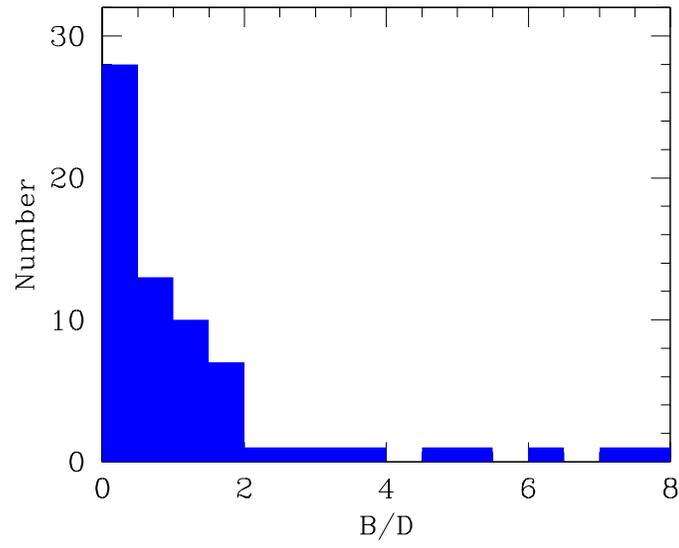}
\caption{ Distribution of bulge-to-disk ratio, B/D, in GOALS galaxies.
}
\end{figure*}
\clearpage

\begin{figure*}[ht]
\includegraphics[width=1.0\textwidth]{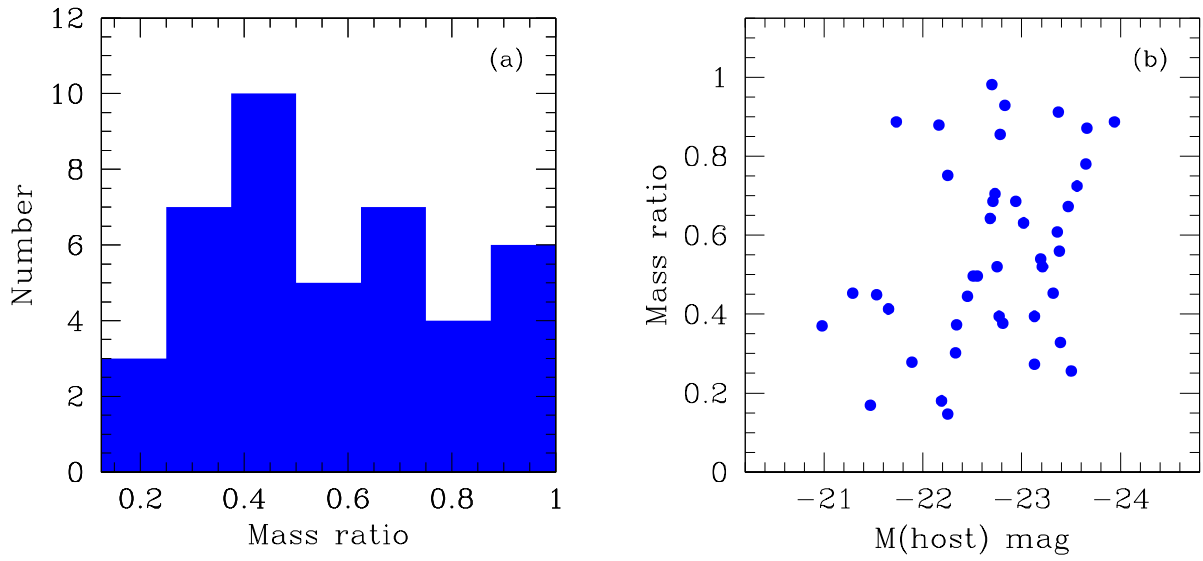}
\caption{ a) Distribution of mass ratio in binary systems and b) mass ratio as a function of M(host).
}
\end{figure*}

\begin{figure*}
\epsscale{1.0}
\includegraphics[width=1.0\textwidth]{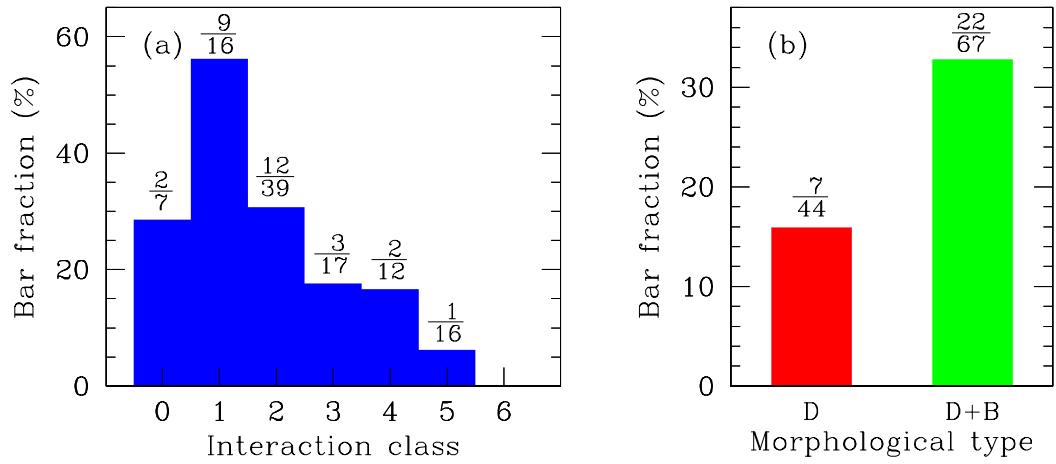}
\caption{Bar fractions as a function of a) interaction class, and b) morphological type.}
\end{figure*}

\begin{figure*}
\epsscale{1.0}
\includegraphics[width=1.0\textwidth]{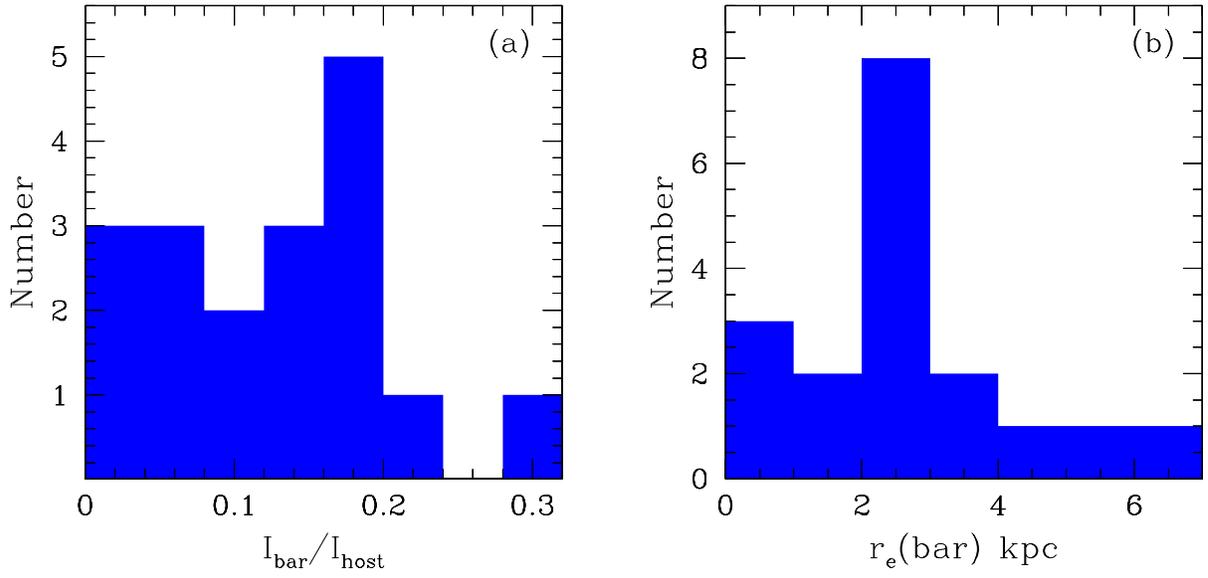}
\caption{a) Distribution of I$_{bar}$/I$_{host}$, and b) bar effective radius r$_e$(bar). }
\end{figure*}
\clearpage

\begin{figure*}
\epsscale{1.0}
\includegraphics[width=1.0\textwidth]{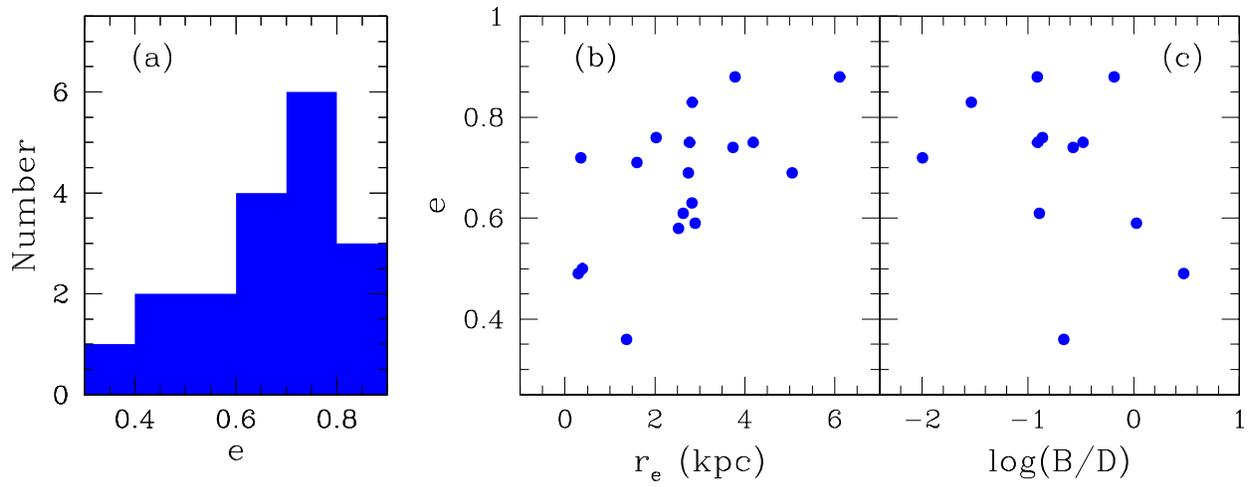}
\caption{a) Distribution of bar ellipticity, bar ellipticity as a function of b) bar half-light radius, and c) bulge-to-disk ratio.}
\end{figure*}
\clearpage

\begin{figure*}[ht]
\includegraphics[width=1.0\textwidth]{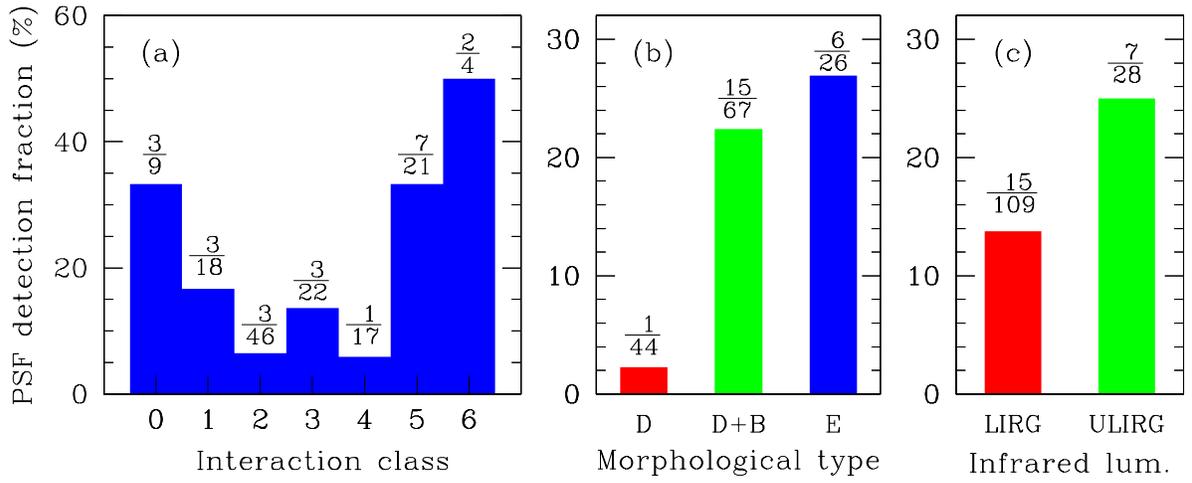}
\caption{PSF detection fractions as a function of a) interaction class, b) morphological type, and c) infrared luminosity (LIRG vs. ULIRG).}
\end{figure*}

\begin{figure*}
\includegraphics[width=1.0\textwidth]{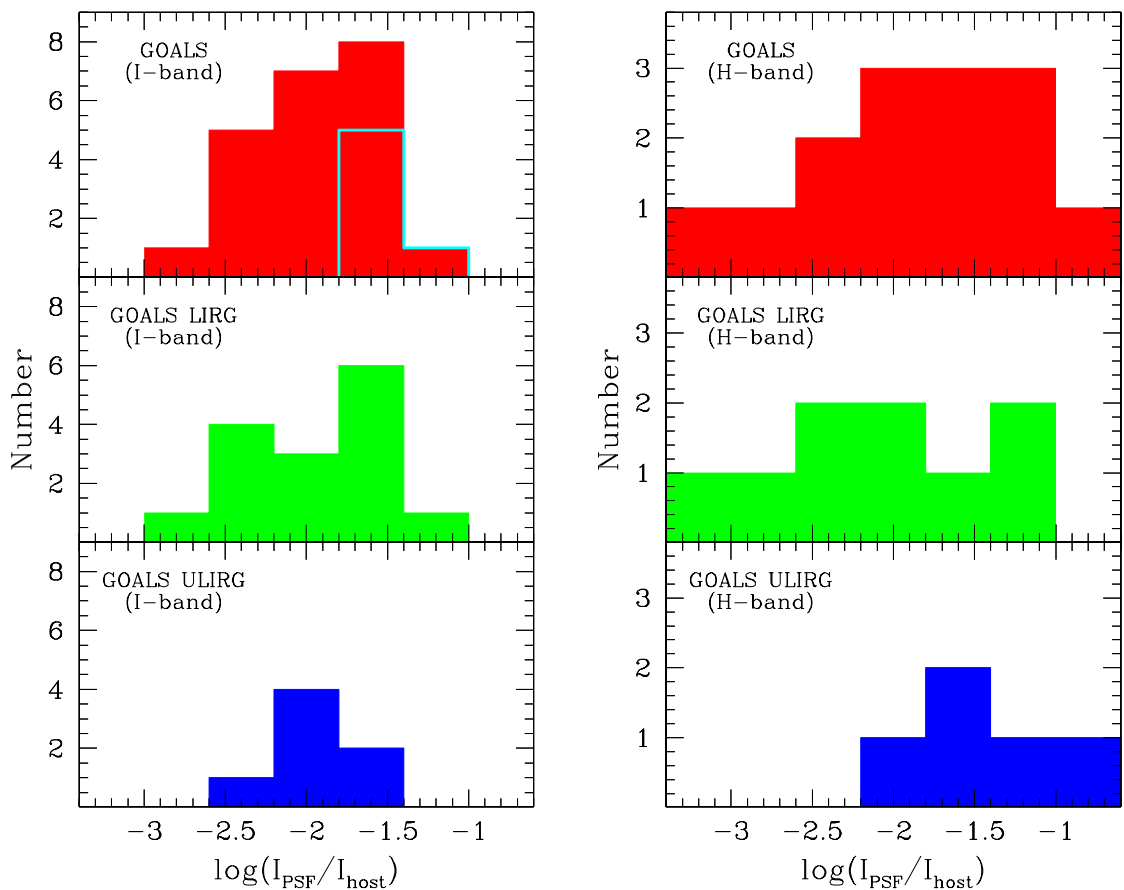}
\caption{ Distribution of I$_{PSF}$/I$_{host}$ in GOALS galaxies (red), GOALS LIRGs (green), and GOALS ULIRGs (blue).
The I$_{PSF}$/I$_{host}$ derived from I-band images is plotted on the left panel, 
and the ratio derived from H-band images for the GOALS galaxies classified as ellipticals is plotted on the right panel.
The cyan line on top-left panel represents I$_{PSF}$/I$_{host}$ in GOALS elliptical galaxies.
}
\end{figure*}

\begin{figure*}[ht]
\includegraphics[width=1.0\textwidth]{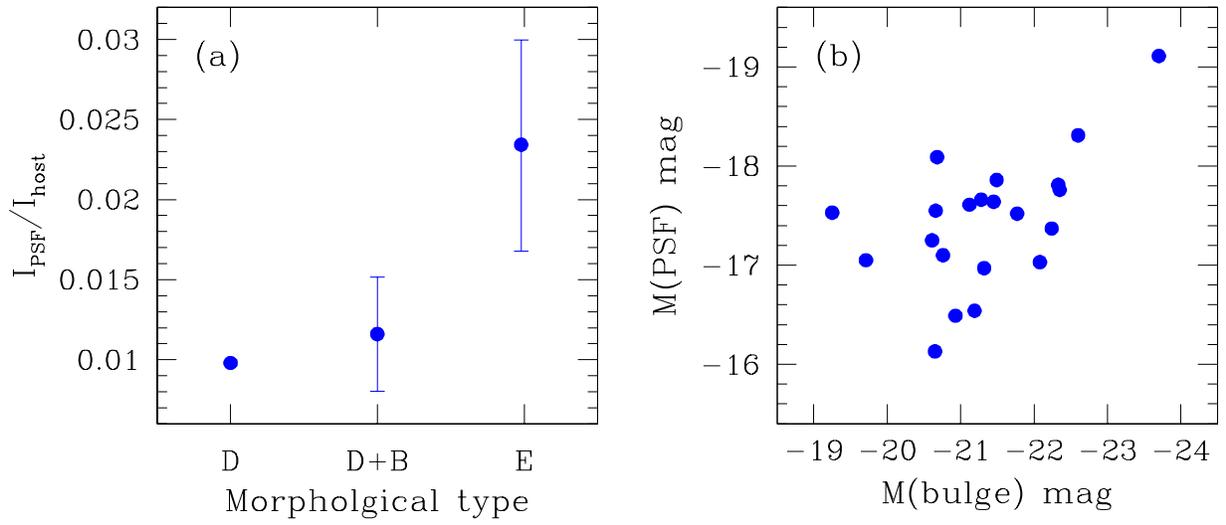}
\caption{a) I$_{PSF}$/I$_{host}$ as a function of morphological type. b) M(bulge) vs. M(PSF) plot.
}
\end{figure*}
\clearpage

\begin{figure*}
\includegraphics[width=1.0\textwidth]{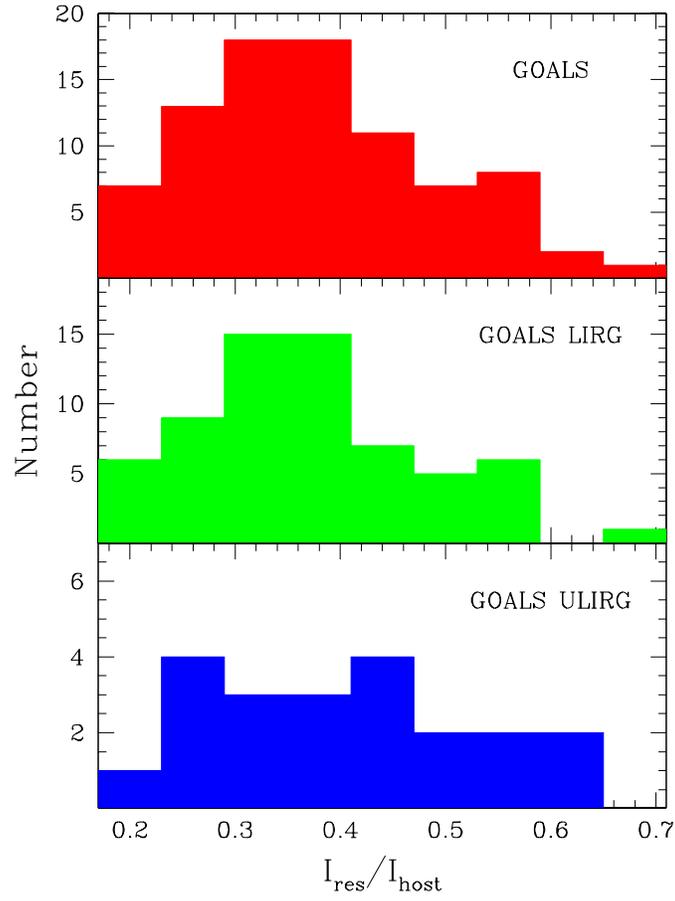}
\caption{ Distribution of I$_{res}$/I$_{host}$ in GOALS galaxies (red), GOALS LIRGs (green), and GOALS ULIRGs (blue).
}
\end{figure*}

\begin{figure*}[ht]
\includegraphics[width=1.0\textwidth]{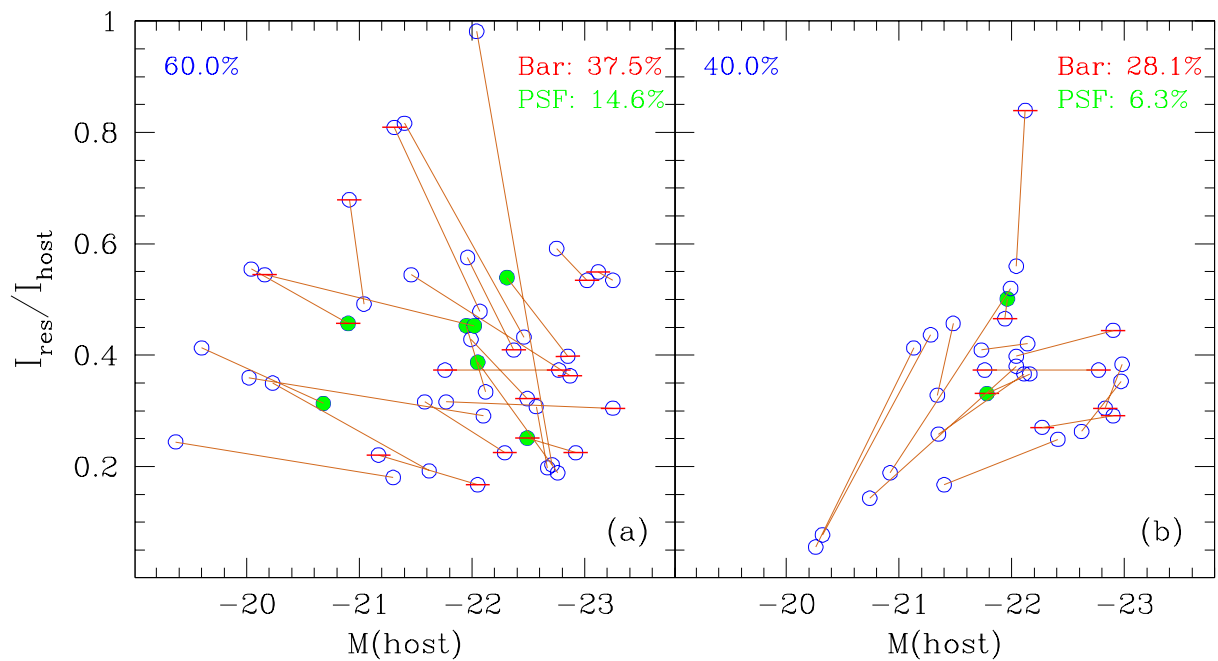}
\caption{Host absolute magnitude vs. I$_{res}$/I$_{host}$ for binary systems. 
In a binary system, if the I$_{res}$/I$_{host}$ in a small companion (faint component) is larger than that in large one (bright component), 
it is plotted on the left panel (a) and the opposite case is plotted on the right panel (b).
In both plots, the interacting pair is connected with a line, and green filled circles and red bars represent
galaxies with PSF and stellar bar components, respectively.
}
\end{figure*}
\clearpage

\begin{figure*}[ht]
\includegraphics[width=1.0\textwidth]{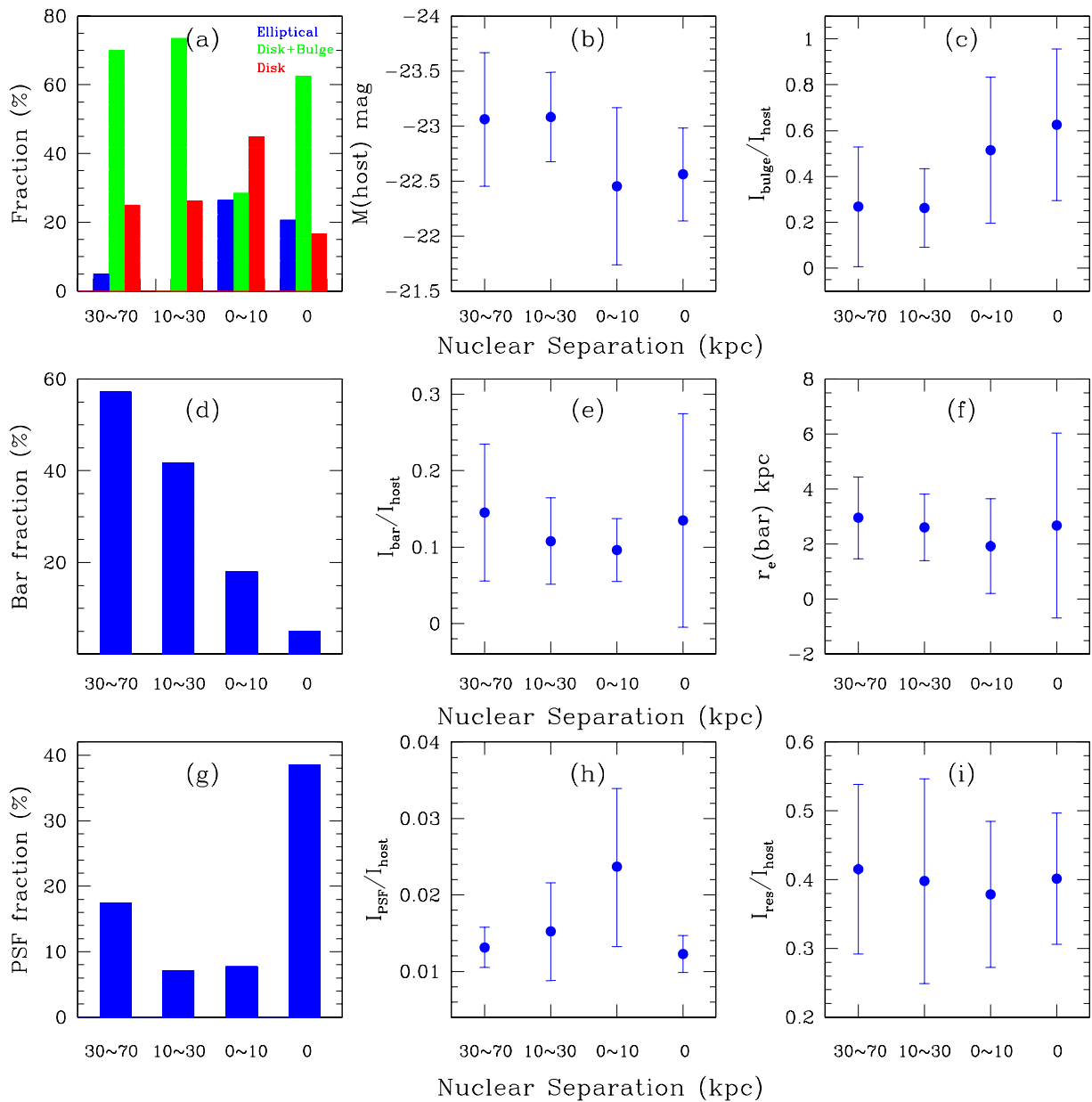}
\caption{ a) Fraction of morphological types, b) host magnitude, c) I$_{bulge}$/I$_{host}$, d) bar fraction, e) I$_{bar}$/I$_{host}$, f) bar size, g) PSF fraction, h) I$_{PSF}$/I$_{host}$, and i) I$_{res}$/I$_{host}$ as a function of nuclear separation.
}
\end{figure*}
\clearpage

\begin{figure*}[ht]
\epsscale{1.0}
\includegraphics[width=1.0\textwidth]{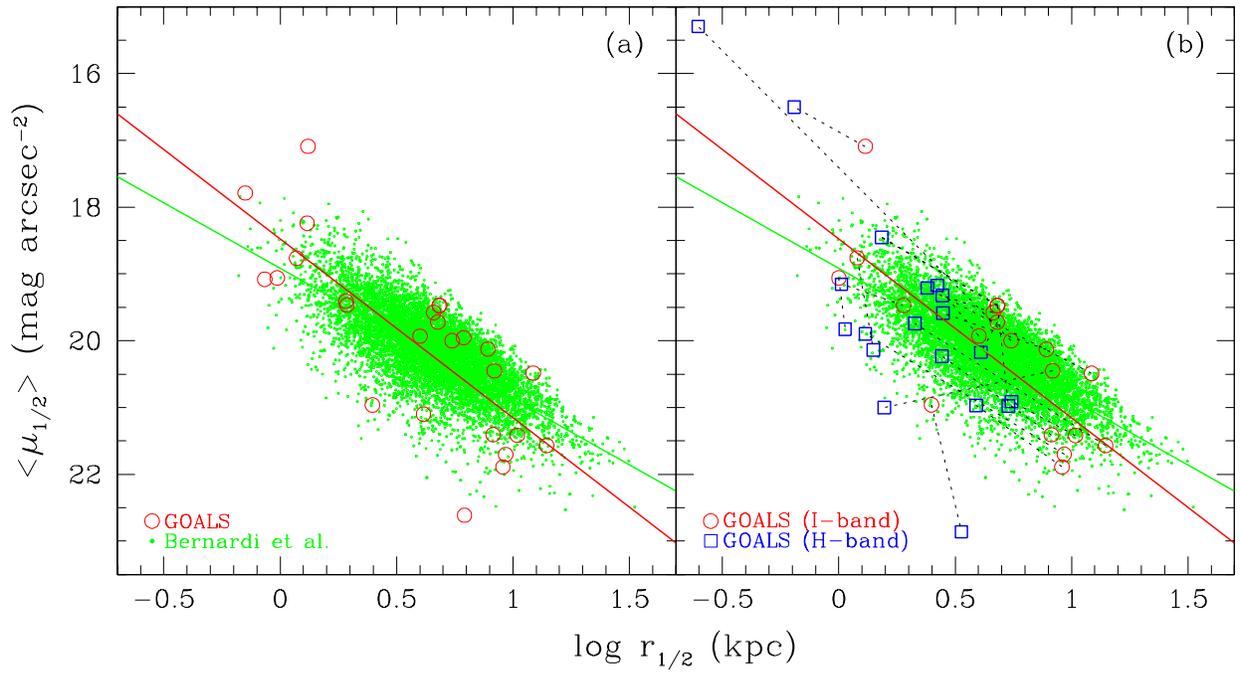}
\caption{Surface brightness vs. half-light radius in the I-band for elliptical hosts in GOALS galaxies (red circles).
The green and red lines represent least-square fits of the SDSS ellipticals and GOALS galaxies, respectively.
The dotted lines in Fig. 16b connect the same objects observed in I-band (circles) and H-band (squares).}
\end{figure*}
\clearpage

\begin{figure*}[ht]
\epsscale{1.0}
\includegraphics[width=1.0\textwidth]{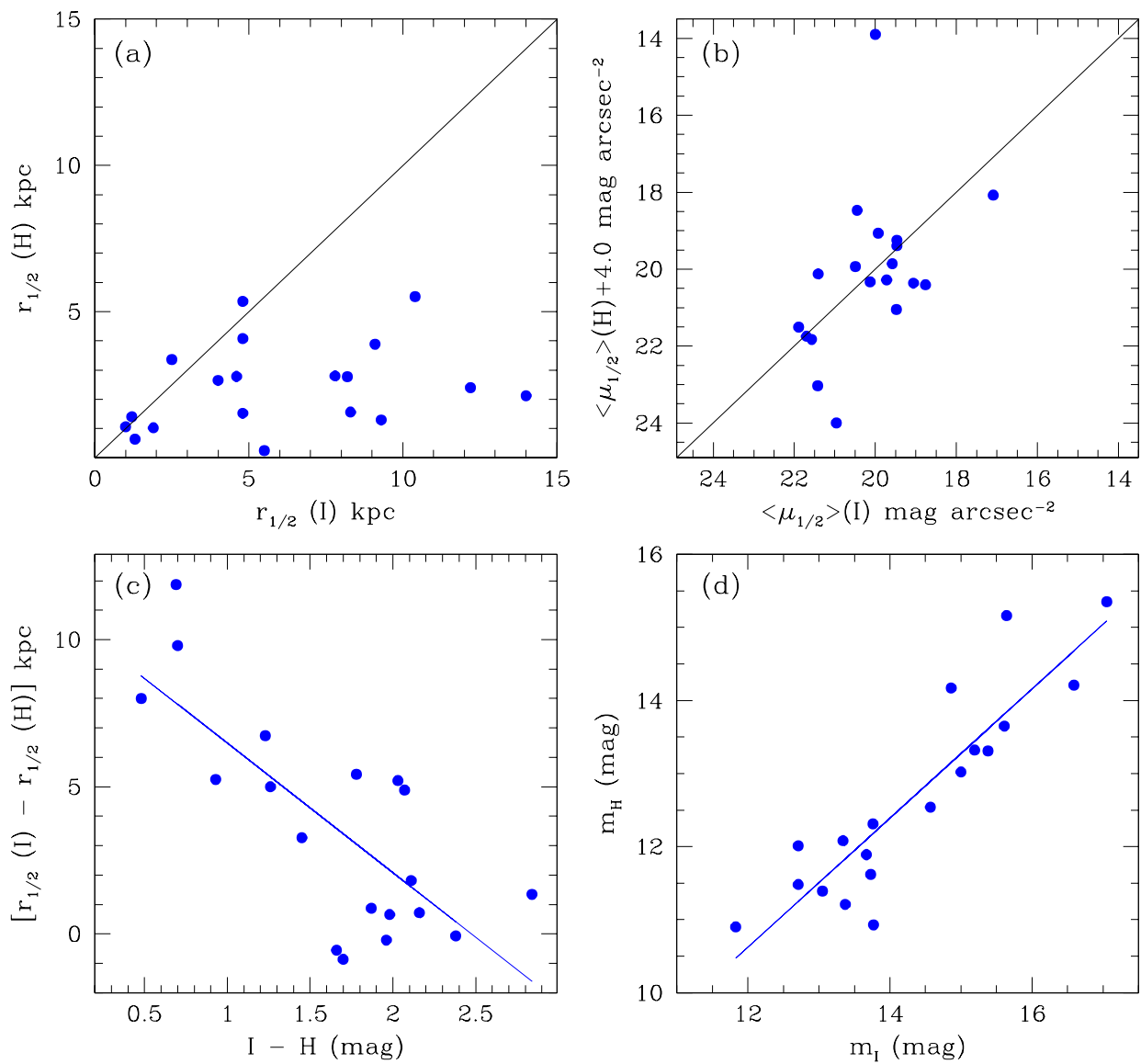}
\caption{ a) Half-light radius and b) surface brightness for GOALS elliptical hosts in I and H bands, 
c) I-H color vs. r$_{1/2}$(I)$-$r$_{1/2}$(H), and d) host magnitude of m$_I$ vs. m$_H$ plots.}
\end{figure*}
\clearpage

\begin{figure*}[ht]
\includegraphics[width=1.0\textwidth]{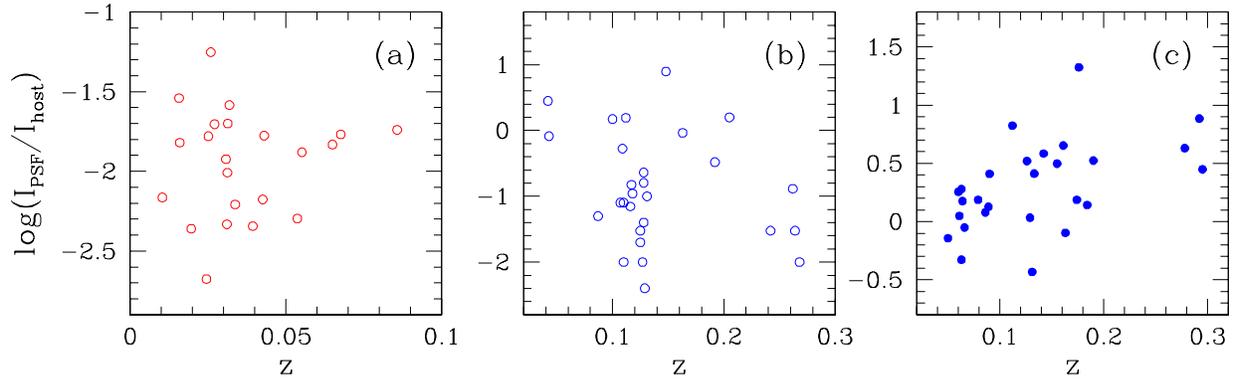}
\caption{ PSF to host intensity ratio I$_{PSF}$/I$_{host}$ as a function of redshift for a) GOALS galaxies, b) QUEST ULIRGs, and c) PG QSOs.
}
\end{figure*}
\clearpage

\begin{figure*}
\epsscale{1.0}
\includegraphics[width=1.0\textwidth]{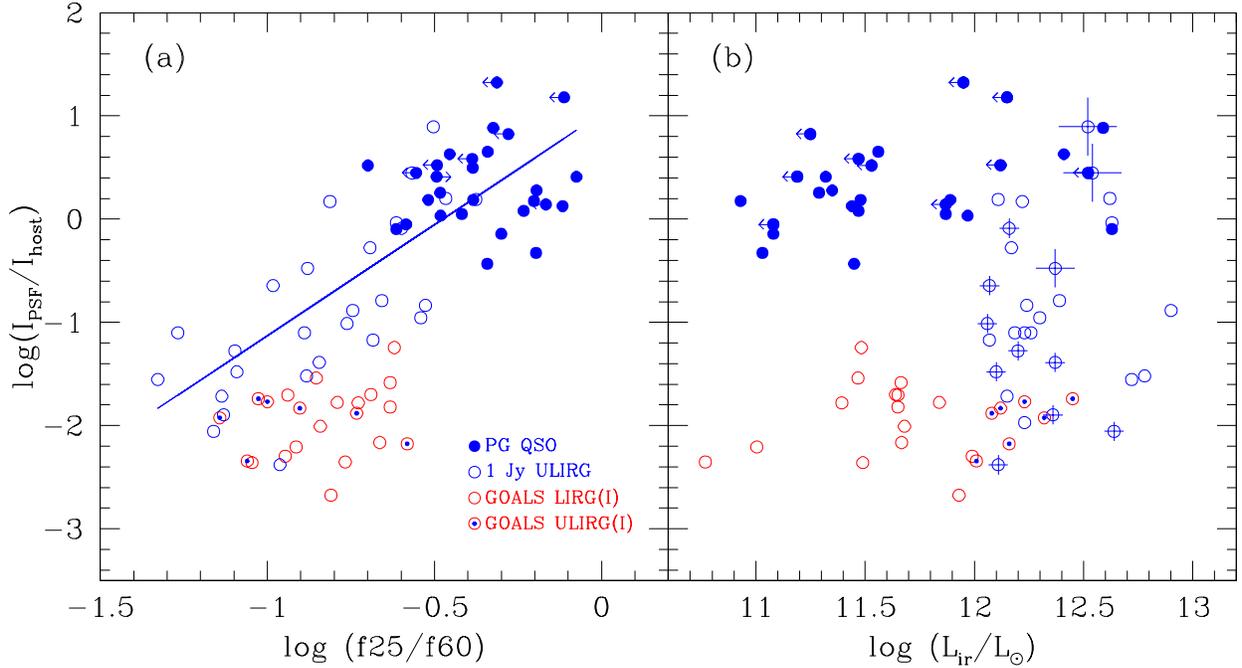}
\caption{I$_{PSF}$/I$_{host}$ as a function a) $f_{25}/f_{60}$, and b) $L_{\rm IR}$.
The red circles, red circles with blue dot, blue circles, and blue filled circles represent
GOALS LIRGs (I-band), GOALS ULIRGs (I-band), QUEST ULIRGs (H-band), and PG QSO hosts (H-band), respectively.
The arrows represent either upper or lower limits and circles with small, medium, and large crosses represent
QUEST ULIRGs with outflow velocity less than 1,000 km s$^{-1}$, 1,000 to 5,000 km s$^{-1}$, and 5,000 to 10,000 km s$^{-1}$ respectively.}
\end{figure*}

\clearpage

\end{document}